\documentclass[12pt,aps,prd,preprint,tightenlines,superscriptaddress,
   showpacs,nofootinbib]{revtex4}
\newcommand{\PRE}[1]{{#1}} 

\usepackage{bm} \usepackage{epsfig}

\newcommand{\mweak}{M_{\text{weak}}}
\newcommand{\mplanck}{M_{\text{Pl}}} 
\newcommand{\mstar}{M_{*}}

\newcommand{\ifb}{\text{fb}^{-1}}

\newcommand{\mev}{\text{MeV}}
\newcommand{\gev}{\text{GeV}} 
\newcommand{\tev}{\text{TeV}}

\newcommand{\s}{\text{s}}

\newcommand{\etal}{{\em et al.}}

\newcommand{\eqref}[1]{Eq.~(\ref{#1})}
\newcommand{\eqsref}[2]{Eqs.~(\ref{#1}) and (\ref{#2})}
\newcommand{\secref}[1]{Sec.~\ref{sec:#1}}

\newcommand{\figref}[1]{Fig.~\ref{fig:#1}}
\newcommand{\figsref}[2]{Figs.~\ref{fig:#1} and \ref{fig:#2}}

\newcommand{\NLSP}{\text{NLSP}}

\newcommand{\mNLSP}{m_{\NLSP}}
\newcommand{\mchi}{m_{\chi}}
\newcommand{\mgravitino}{m_{\gravitino}}
\newcommand{\mmed}{M_{\text{med}}}
\newcommand{\gravitino}{\tilde{G}} 
\newcommand{\Bino}{\tilde{B}}
 
\newcommand{\stau}{\tilde{\tau}}
\newcommand{\snu}{\tilde{\nu}}

\newcommand{\bold}[1]{{\text{\normalsize\bm{$#1$}}}}
\newcommand{\rem}[1]{{}}

\begin{document}

\preprint{UCI-TR-2004-11}  \preprint{hep-ph/0404231}

\title{
\PRE{\vspace*{1.5in}}
Supergravity with a Gravitino LSP
\PRE{\vspace*{0.3in}}
}

\author{Jonathan L.~Feng}
\affiliation{Department of Physics and Astronomy, University of
California, Irvine, CA 92697, USA
\PRE{\vspace*{.1in}}
}
\author{Shufang Su}
\affiliation{Department of Physics, University of Arizona, Tucson, AZ
85721, USA
\PRE{\vspace*{.5in}}
}

\author{Fumihiro Takayama%
\PRE{\vspace*{.2in}}
} 
\affiliation{Department of Physics and Astronomy, University of
California, Irvine, CA 92697, USA
\PRE{\vspace*{.1in}}
}


\begin{abstract}
\PRE{\vspace*{.3in}} We investigate supergravity models in which the
lightest supersymmetric particle (LSP) is a stable gravitino.  We
assume that the next-lightest supersymmetric particle (NLSP) freezes
out with its thermal relic density before decaying to the gravitino at
time $t \sim 10^4 - 10^8~\s$.  In contrast to studies that assume a
fixed gravitino relic density, the thermal relic density assumption
implies upper, not lower, bounds on superpartner masses, with
important implications for particle colliders.  We consider slepton,
sneutrino, and neutralino NLSPs, and determine what superpartner
masses are viable in all of these cases, applying CMB and
electromagnetic and hadronic BBN constraints to the leading two- and
three-body NLSP decays.  Hadronic constraints have been neglected
previously, but we find that they provide the most stringent
constraints in much of the natural parameter space.  We then discuss
the collider phenomenology of supergravity with a gravitino LSP.  We
find that colliders may provide important insights to clarify BBN and
the thermal history of the Universe below temperatures around 10 GeV
and may even provide precise measurements of the gravitino's mass and
couplings.
\end{abstract}


\maketitle

\section{Introduction}
\label{sec:introduction}

Supersymmetric theories predict the existence of a spin 3/2 particle,
the gravitino, the partner of the spin 2 graviton.  The gravitino mass
is 
\begin{equation}
\mgravitino \sim \frac{F}{\mstar} \ ,
\end{equation}
where $F$ is the scale of supersymmetry breaking, and $\mstar = (8 \pi
G_N)^{-1/2} \simeq 2.4 \times 10^{18}~\gev$ is the reduced Planck
mass.  The masses of scalar superpartners are derived from terms such
as
\begin{equation}
\lambda_{ij} \int d^4\theta \frac{Z^\dagger Z \Phi_i^\dagger
  \Phi_j}{\mmed^2} \ ,
\end{equation}
where $\lambda_{ij}$ are unknown constants, $Z$ is a superfield whose
auxiliary component develops the vacuum expectation value $F$,
$\Phi_i$ are standard model superfields, and $\mmed$ is the mass scale
of the interactions that mediate supersymmetry breaking.  Similar
terms give the spin $1/2$ superpartners mass.  In supergravity, the
mediating interactions are gravitational, and so $\mmed \sim \mstar$,
$F \sim (10^{10}~\gev)^2$, and the gravitino and all standard model
superpartners have mass $\sim F/\mstar \sim \mweak$, with the precise
ordering determined by unknown constants, such as $\lambda_{ij}$.

Most studies of supergravity have assumed, either explicitly or
implicitly, that the lightest supersymmetric particle (LSP) is a
standard model superpartner.  This avoids potential complications
resulting from the decay of standard model superpartners to a
gravitino LSP, which naturally happens at time $t \sim 10^4 -
10^8~\s$, well after Big Bang nucleosynthesis (BBN).  However, the
phenomenology and cosmology of gravitinos have also been considered in
a number of studies~\cite{Pagels:ke,Weinberg:zq,Krauss:1983ik,%
Nanopoulos:1983up,Khlopov:pf,Ellis:1984eq,Ellis:1984er,%
Juszkiewicz:gg,Ellis:1990nb,Moroi:1993mb,Bolz:2000fu,Khlopov:rs}.
(See also related studies of axino and quintessino dark
matter~\cite{Covi:1999ty,Covi:2001nw,Bi:2003qa,Hooper:2004qf}.)

More recently, it has been shown that the gravitino LSP possibility
does not destroy the beautiful predictions of BBN even when the
gravitino LSPs produced in late decays have relic density
$\Omega_{\gravitino} = 0.23$ and so are present in sufficient numbers
to account for all of dark
matter~\cite{Feng:2003xh,Feng:2003uy,threebody}.  In fact, bounds from
the cosmic microwave background (CMB) and, in some corners of
parameter space, entropy production and the diffuse photon spectrum
may be even more severe than bounds from
BBN~\cite{Feng:2003xh,Feng:2003uy}.  Nevertheless, all of these bounds
were shown to be respected for some regions of parameter space with
weak scale superpartners.  The possibility of superweakly-interacting
massive particle (superWIMP) gravitino dark matter from NLSP decays
thus appears to be viable.  The analogous scenario in extra
dimensional theories~\cite{Feng:2003xh,Feng:2003uy,Feng:2003nr}, as
well as interesting astrophysical implications in this and related
scenarios~\cite{Bi:2003qa,Chen:2003gz,Sigurdson:2003vy} have also been
discussed.

In this work, we take an approach that differs from the exploration of
superWIMP gravitino dark matter.  Instead of assuming that gravitinos
are the dark matter with $\Omega_{\gravitino} = 0.23$, we assume that
the NLSP reaches its thermal relic density
$\Omega_{\NLSP}^{\text{th}}$ before decaying, and so
$\Omega_{\gravitino} = (m_{\gravitino}/m_{\NLSP} )
\Omega_{\NLSP}^{\text{th}}$.  That is, we relax the constraint that
gravitinos from NLSP decays account for all of dark matter. Rather we
assume the simplest thermal history for the Universe and ask what
regions of $(\mgravitino, \mNLSP)$ parameter space are allowed.  The
thermal relic density assumption has consequences that differ markedly
from the fixed gravitino relic density assumption. To see this, assume
that the gravitino and NLSP masses are both parametrized by a general
superpartner mass scale $m_{\text{SUSY}}$.  As noted in
Ref.~\cite{threebody}, if one assumes a fixed gravitino relic density,
the NLSP number density scales as $1/m_{\text{SUSY}}$.  Low
superpartner masses are therefore disfavored.  In contrast, if one
assumes a thermal relic density for the NLSP,
$\Omega_{\text{NLSP}}^{\text{th}} \propto \langle \sigma v \rangle
^{-1} \propto m_{\text{SUSY}}^2$, where $\langle \sigma v \rangle$ is
the thermally-averaged NLSP annihilation cross section. The NLSP
number density then scales as $m_{\text{SUSY}}$, and so high
superpartner masses are disfavored.  This difference has obviously
important implications for collider searches for new physics, and we
discuss collider implications below.

Even given the NLSP thermal relic density assumption, gravitinos may
still be all of dark matter --- for example, if the gravitino relic
density from NLSP decays is too low, the remainder may be made up by
gravitinos produced during reheating.  However, the existence of such
alternative gravitino sources is either untestable, or testable only
with strong assumptions about the early Universe.  In contrast, the
existence of a gravitino component from NLSP decays makes several
robust predictions that are testable at cosmological observatories and
collider experiments, and we concentrate on this gravitino source
here.  Before leaving the topic of reheating altogether, however, we
note that the gravitino LSP scenario has an important virtue with
respect to reheating.  For stable weak-scale gravitinos, the
overclosure constraint is well-known to require a bound on reheat
temperature of $T_R \alt 10^{10}~\gev$~\cite{Bolz:2000fu}.  Recently,
however, it has been shown that if the gravitino is not the LSP,
hadronic BBN constraints greatly strengthen this
bound~\cite{Kawasaki:2004yh}.  For example, if the gravitino decays to
the LSP $+$ hadrons with branching fraction $10^{-3}$, the reheat
temperature must satisfy $T_R \alt 10^6~\gev$.  This is uncomfortably
low.  The gravitino LSP scenario is therefore preferred if one
requires a high reheat temperature, as might be desirable, for
example, for scenarios of leptogenesis~\cite{Fujii:2003nr}.

In the present analysis, in addition to the constraints on
electromagnetic energy release considered previously, we include the
recent results on hadronic BBN constraints~\cite{Kawasaki:2004yh}.
The work of Ref.~\cite{Kawasaki:2004yh} represents a significant
update to previous hadronic analyses~\cite{Reno:1987qw,%
Dimopoulos:1987fz,Dimopoulos:1988ue,Kohri:2001jx}.  To include these
results correctly, we must, of course, determine the leading
contributions to hadronic energy.  For slepton\footnote{Throughout
this work, ``slepton'' refers to a charged slepton.} and sneutrino
NLSPs, the leading contribution is from three-body decays
\begin{eqnarray}
\tilde{l} &\to& l Z \tilde{G} \ , \ \nu W \tilde{G} \nonumber \\
\tilde{\nu} &\to& \nu Z \tilde{G} \ , \ l W \tilde{G} \ .
\label{threebody}
\end{eqnarray}
The three-body decays have been studied in Ref.~\cite{threebody}.  For
a neutralino NLSP, the leading contribution to hadronic energy is from
the two-body decays, such as
\begin{equation}
\chi \to Z \gravitino , \, h \gravitino\ ,
\label{twobody}
\end{equation}
followed by $Z, h \to q \bar{q}$.  These decays, and the hadronic
constraints on them, were neglected in previous works.  As we will
see, however, they are the leading constraints in much of parameter
space and they are especially important when the superpartner masses
and their splittings are all of the order of the weak scale, the most
natural possibility.  

The decays of \eqsref{threebody}{twobody} may be suppressed
kinematically if $m_{\text{NLSP}} - \mgravitino < m_Z, m_W$ or
dynamically, as when the neutralino is photino-like.  However, even in
these cases, decays such as $\tilde{l} \to l q \bar{q} \gravitino$ and
$\chi \to q \bar{q} \gravitino$ are still possible at higher order.
We have included estimates of these in our analysis.  These decays are
in some sense ``model-independent''; even in the extreme case where
the dominant decay is to invisible particles, at higher order there
will be contributions to hadronic cascades from such decays.  The
hadronic bounds are so constraining that these should be considered
for any late decaying particle, whether a superpartner, an axino, a
modulus, or other particle.

After determining the regions of parameter space allowed by cosmology,
we discuss the collider signals.  The upper bounds on superpartner
masses resulting from the thermal relic density assumption imply
promising prospects for superpartners to be within reach of future
collider experiments.  In addition, we will see that the signals of
supersymmetry in gravitino LSP scenarios may be completely different
{}from the conventional supersymmetry signals.  In particular, if
sufficient NLSPs can be collected and monitored for decays, the NLSP
lifetime may be measured, which may considerably sharpen our
understanding of BBN and the thermal history of the Universe at
temperatures of 10 GeV and below.  Such studies may also provide the
first direct measurements of the gravitino mass and the Planck scale
from particle physics~\cite{Buchmuller:2004rq,Feng:2004gn}.

The gravitino LSP possibility, assuming a thermal NLSP relic density,
has been discussed recently in the context of minimal
supergravity~\cite{Ellis:2003dn}.  Our work is complementary in that
we do not work in a specific model framework, but rather consider
several NLSP candidates, as well as a wide range of gravitino and NLSP
masses.  Our work also differs in that we consider the hadronic
constraints and the leading two- and three-body decays that contribute
to hadronic energy.  As noted above, we find that these are the
leading constraints in the most natural regions of parameter space.

\section{Late Decays}
\label{sec:late}

We first discuss the decays of NLSPs for each of the various NLSP
cases.  NLSPs freeze out and are highly non-relativistic when they
decay.  We will be most interested in deriving the electromagnetic
(EM) and hadronic energy releases
\begin{eqnarray}
\xi_{\text{i}} \equiv 
\epsilon_{\text{i}} B_{\text{i}} Y_{\text{NLSP}} \ ,
\label{xi}
\end{eqnarray}
where $\text{i} = \text{EM}, \text{had}$, because BBN constraints are,
to a good approximation, constraints on $\xi_{\text{EM}}$ and
$\xi_{\text{had}}$.  Here $B_{\text{i}}$ is the branching fraction
into EM/hadronic components, and $\epsilon_{\text{i}}$ is the
EM/hadronic energy released in each NLSP decay.  These are discussed
in this section.  $Y_{\text{NLSP}} \equiv
n_{\text{NLSP}}/n_{\gamma}^{\text{BG}}$ is the NLSP number density
just before NLSP decay, normalized to the background photon number
density $n_{\gamma}^{\text{BG}} = 2 \zeta(3) T^3 /
\pi^2$.\footnote{Another common definition is $Y_{\text{NLSP}} \equiv
n_{\text{NLSP}}/s$, where $s = (2 \pi^2/45) g_{* S} T^3$ is the
entropy density.  In the era of NLSP decays to gravitinos, $s \simeq
7.0 \, n_{\gamma}^{\text{BG}}$.}  Given the assumptions of this work,
$Y_{\text{NLSP}}$ is determined by the thermal relic density for each
NLSP; it is discussed in \secref{thermal}.

We will consider the cases of slepton, sneutrino, and neutralino
NLSPs.  As specific examples in each of these categories, we will
focus on $\stau_R$, $\snu_{\tau}$, and $\Bino$ NLSPs, but our results
are easily extended to the general cases.

\subsection{Slepton NLSP}

\begin{figure}
\resizebox{3.25 in}{!}{
\includegraphics{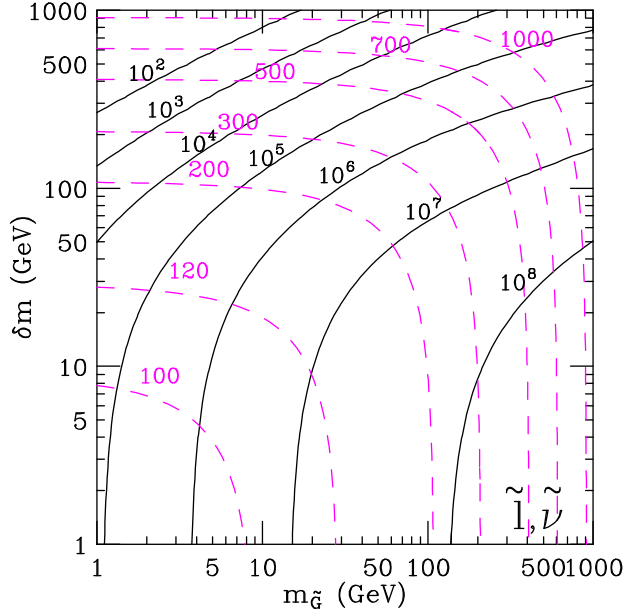}}
\caption{NLSP lifetime in seconds (solid) and mass in GeV (dashed) in
the $(m_{\gravitino}, \delta m \equiv m_{\text{NLSP}} - m_{\gravitino}
- m_Z)$ plane for slepton and sneutrino NLSPs.
\label{fig:life_slep} }
\end{figure}

The width for the decay of any sfermion to a gravitino is
\begin{equation}
 \Gamma(\tilde{f} \to f \tilde{G}) =\frac{1}{48\pi \mstar^2}
 \frac{m_{\tilde{f}}^5}{m_{\tilde{G}}^2} 
 \left[1 -\frac{m_{\tilde{G}}^2}{m_{\tilde{f}}^2} \right]^4 \ ,
\label{sfermionwidth}
\end{equation}
assuming the fermion mass is negligible.  For $\Delta m \equiv
m_{\tilde{f}} - m_{\gravitino} \ll m_{\gravitino}$, the sfermion decay
lifetime is
\begin{eqnarray}
 \tau(\tilde{f} \to f \tilde{G})
\simeq 3.6\times 10^8~\s
\left[\frac{100~\gev}{\Delta m}\right]^4
\left[\frac{m_{\tilde{G}}}{1~\tev}\right]\ .
\label{sfermionlifetime}
\end{eqnarray}
The slepton lifetime and mass are given in the $(m_{\gravitino},
\delta m)$ plane in \figref{life_slep}, where we have defined
\begin{equation}
\delta m \equiv \Delta m - m_Z = m_{\text{NLSP}} - m_{\gravitino} -
m_Z \ ,
\end{equation}
a useful measure of the kinematically available energy in three-body
decays to be discussed below.

For selectrons, the produced electron in these two-body decays
immediately initiates an electromagnetic (EM) cascade, and so
\begin{equation}
B_{\text{EM}}^{\tilde{e}} \simeq 1 \ , \quad 
\epsilon_{\text{EM}}^{\tilde{e}} 
= \frac{m_{\tilde{e}}^2 - m_{\gravitino}^2} {2 m_{\tilde{e}}} \ .
\end{equation}
For smuons, the produced muon typically interacts with the background
photons before decaying~\cite{Feng:2003uy}, and so
\begin{equation}
B_{\text{EM}}^{\tilde{\mu}} \simeq 1 \ , \quad 
\epsilon_{\text{EM}}^{\tilde{\mu}} 
= \frac{m_{\tilde{\mu}}^2 - m_{\gravitino}^2} {2 m_{\tilde{\mu}}} \ .
\end{equation}
For staus, the resulting $\tau$ sometimes decays into mesons, which
could in principle induce hadronic cascades.  As shown in
Refs.~\cite{Feng:2003uy,threebody}, however, for decay times $\tau >
10^3 - 10^4~\s$, the hadronic interaction time of all pions and kaons
is much longer than their decay time.  The decays of staus therefore
typically contribute only to EM cascades, and we assume this in the
following analysis.  In contrast to the selectron and smuon cases,
however, on average, about half of the $\tau$ energy is lost to
neutrinos.  We therefore have
\begin{equation}
B_{\text{EM}}^{\tilde{\tau}} \simeq 1 \ , \quad 
\epsilon_{\text{EM}}^{\tilde{\tau}} \approx \frac{1}{2}
\frac{m_{\tilde{\tau}}^2 - m_{\gravitino}^2} {2 m_{\tilde{\tau}}} \ .
\end{equation}

As noted in \secref{introduction}, three-body decays are also
important when they are the leading contribution to hadronic cascades.
They are therefore important for slepton NLSPs.  The decays are those
of \eqref{threebody}.  The decay $\tilde{l} \to l Z \gravitino$ takes
place through off-shell $l$, $\tilde{l}$, and $\chi$, and also through
a four-point interaction.  The three-body decay widths for sleptons
have been discussed and presented in Ref.~\cite{threebody}, 
and we refer readers there for details. Given these decay widths, the
hadronic branching fraction is
\begin{equation}
B_{\text{had}}^{\tilde{l}} \simeq
\frac{\Gamma(\tilde{l}\to l Z \tilde{G}) B_{\text{had}}^Z
+\Gamma(\tilde{l}\to \nu W \tilde{G})B_{\text{had}}^W
+\Gamma(\tilde{l}\to l q \bar{q} \tilde{G})}
{\Gamma(\tilde{l}\to l \tilde{G})} \ ,
\end{equation}
for $\tilde{l} = \tilde{e}$, $\tilde{\mu}$, $\tilde{\tau}$, where
$B_{\text{had}}^Z$,$B_{\text{had}}^W \approx 0.7$ are the $Z$ and $W$
hadronic branching fractions.  $\Gamma(\tilde{l}\to \nu W
\tilde{G})=0$ for purely right-handed sleptons.  Below, we will
consider cases in which the three-body decays are kinematically
allowed.  These decay modes may nevertheless become suppressed for
$\Delta m \sim m_Z, m_W$.  However, even for such small mass
splittings, hadronic decays are still possible through higher order
decays.  With this in mind, we have included the four-body process
$\Gamma(\tilde{l}\to l q \bar{q} \tilde{G})$.  We have not calculated
this width.  However, we expect $B ( \tilde{l}\to l q \bar{q}
\tilde{G}) \sim 10^{-6}$, and we take this value, which provides a
lower limit on $B_{\text{had}}^{\tilde{l}}$.

$B_{\text{had}}^{\tilde{l}}$ is typically in the range $10^{-2} -
10^{-5}$, depending on the underlying scale and mass splitting.  As
the branching fraction may vary over a few orders of magnitude,
variations in $\epsilon_{\text{had}}^{\tilde{l}}$ are subdominant.  We
therefore take simply
\begin{equation}
\epsilon_{\text{had}}^{\tilde{l}} 
= \frac{1}{3} \left(m_{\tilde{l}}-m_{\gravitino} \right)
\end{equation}
in our analysis.

\subsection{Sneutrino NLSP}
\label{sec:sneutrino}

The decay width and time for $\tilde{\nu} \to \nu \gravitino$ are
given in \eqsref{sfermionwidth}{sfermionlifetime}, and plotted in
\figref{life_slep}.  These two-body decays are essentially invisible
and do not contribute to either EM or hadronic cascades.  (We neglect
the effects of neutrino thermalization through processes like $\nu
e_{\text{BG}} \to \nu e$.) The three-body decays are therefore even
more important for sneutrinos than sleptons.  These decays have also
been discussed and presented in Ref.~\cite{threebody}.  For
sneutrinos, we have
\begin{eqnarray}
B_{\text{EM}}^{\tilde{\nu}} &\simeq& 
\frac{\Gamma(\tilde{\nu}\to \nu Z \tilde{G})
+\Gamma(\tilde{\nu}\to l W \tilde{G})
+\Gamma(\tilde{\nu}\to \nu f \bar{f} \tilde{G})}
{\Gamma(\tilde{\nu}\to \nu \tilde{G})} \\
\epsilon_{\text{EM}}^{\tilde{\nu}} &=& 
\frac{1}{3} \left(m_{\tilde{\nu}}-m_{\tilde{G}} \right) \\
B_{\text{had}}^{\tilde{\nu}} &\simeq& 
\frac{\Gamma(\tilde{\nu}\to \nu Z \tilde{G}) B_{\text{had}}^Z
+\Gamma(\tilde{\nu}\to l W \tilde{G})B_{\text{had}}^W
+\Gamma(\tilde{\nu}\to \nu q \bar{q} \tilde{G})}
{\Gamma(\tilde{\nu}\to \nu \tilde{G})} \\
\epsilon_{\text{had}}^{\tilde{\nu}} &=& 
\frac{1}{3} \left(m_{\tilde{\nu}}-m_{\tilde{G}} \right) \ ,
\end{eqnarray}
for $\tilde{\nu} = \tilde{\nu}_e$, $\tilde{\nu}_{\mu}$,
$\tilde{\nu}_{\tau}$.  The EM branching fraction is in fact slightly
reduced by decays $\tilde{\nu} \to \nu Z \gravitino$ followed by $Z
\to \nu \bar{\nu}$.  We have neglected this effect.  The four-body
decay takes place through virtual neutralinos.  We again assume $B (
\tilde{\nu}\to \nu q \bar{q} \tilde{G}) \sim 10^{-6}$, which provides
a lower limit on $B_{\text{had}}^{\tilde{\nu}}$.

Note that, in contrast to the case of the slepton NLSP, the EM
branching fraction is suppressed and of the same order as the hadronic
branching fraction.  In our analysis below we have included the EM
constraint, but we find that it is so weak that it does not disfavor
any of the parameter space appearing in figures below.  The hadronic
BBN constraint is so much stronger than the EM constraint at early
times, however, that it is still important then, as we will see.

\subsection{Neutralino NLSP}

For neutralino NLSPs, the decay width to photons is
\begin{equation}
\Gamma(\chi \to \gamma \gravitino) 
= 
\frac{| \bold{N}_{11} \cos \theta_W + \bold{N}_{12} \sin \theta_W |^2}
{48\pi M_*^2} \ 
\frac{m_{\chi}^5}{m_{\gravitino}^2} 
\left[1 - \frac{m_{\gravitino}^2}{m_{\chi}^2} \right]^3 
\left[1 + 3 \frac{m_{\gravitino}^2}{m_{\chi}^2} \right] \ ,
\label{neutralinogamma}
\end{equation}
where $\chi \equiv \bold{N}_{11} (-i \tilde{B}) + \bold{N}_{12} (-i
\tilde{W}) + \bold{N}_{13} \tilde{H}_d + \bold{N}_{14} \tilde{H}_u$.
In the limit $\Delta m \ll m_{\chi}$, the decay lifetime is
\begin{equation}
\tau(\chi \to \gamma \gravitino) 
\approx 2.3 \times 10^7~\s \,
\frac{\cos^2\theta_W}
{| \bold{N}_{11} \cos \theta_W + \bold{N}_{12} \sin \theta_W |^2}
\left[ \frac{100~\gev}{\Delta m} \right]^3  \ ,
\label{neutralinolifetime}
\end{equation}
proportional to $(\Delta m)^3$ and independent of the overall
superpartner mass scale.

This decay contributes only to EM energy.  As noted in
\secref{introduction}, the leading contribution to hadronic energy is
from $\chi \to Z \gravitino,\, h \gravitino$.  These decays produce EM
energy for all possible $Z$ and $h$ decay modes (except $Z \to \nu
\bar{\nu}$), but they may also produce hadronic energy when followed
by $Z, h \to q \bar{q}$.  The decay width to $Z$ bosons
is\footnote{Our expression for $G$ in \eqref{G} differs from the
result of Ref.~\cite{Ellis:2003dn} in the sign of ``12'' in the second
term.  The authors of Ref.~\cite{Ellis:2003dn} used Eq.~(4.31) of
Ref.~\cite{Moroi:1995fs}, which contains a sign error.  We have
corrected for this error.  We thank Y.~Santoso and T.~Moroi for
helpful correspondence.}
\begin{eqnarray}
\Gamma (\chi \to Z \gravitino) &=&
\frac{| - \bold{N}_{11} \sin \theta_W + \bold{N}_{12} \cos \theta_W |^2}
{48\pi M_*^2} \frac{\mchi^5}{\mgravitino^2}
F(\mchi,\mgravitino,m_Z) \nonumber \\
&& \times \left[ \left( 1-\frac{\mgravitino^2}{\mchi^2} \right)^2
\left( 1 + 3 \frac{\mgravitino^2}{\mchi^2} \right)
-\frac{m_Z^2}{\mchi^2}  G (\mchi,\mgravitino,m_Z) \right] \ ,
\end{eqnarray}
where 
\begin{eqnarray}
F (\mchi,\mgravitino,m_Z) &=& \left[ 
\left(1- \left( \frac{\mgravitino + m_Z}{\mchi} \right)^2 \right) 
\left(1- \left( \frac{\mgravitino - m_Z}{\mchi} \right)^2 \right) 
\right]^{1/2} \label{F} \\
G (\mchi,\mgravitino,m_Z) &=& 3 + \frac{\mgravitino^3}{\mchi^3} 
\left(-12 + \frac{\mgravitino}{\mchi} \right)
 + \frac{m_Z^4}{\mchi^4} - \frac{m_Z^2}{\mchi^2}
\left(3 - \frac{\mgravitino^2}{\mchi^2} \right) \ .
\label{G}
\end{eqnarray}

\begin{figure}
\resizebox{3.25 in}{!}{
\includegraphics{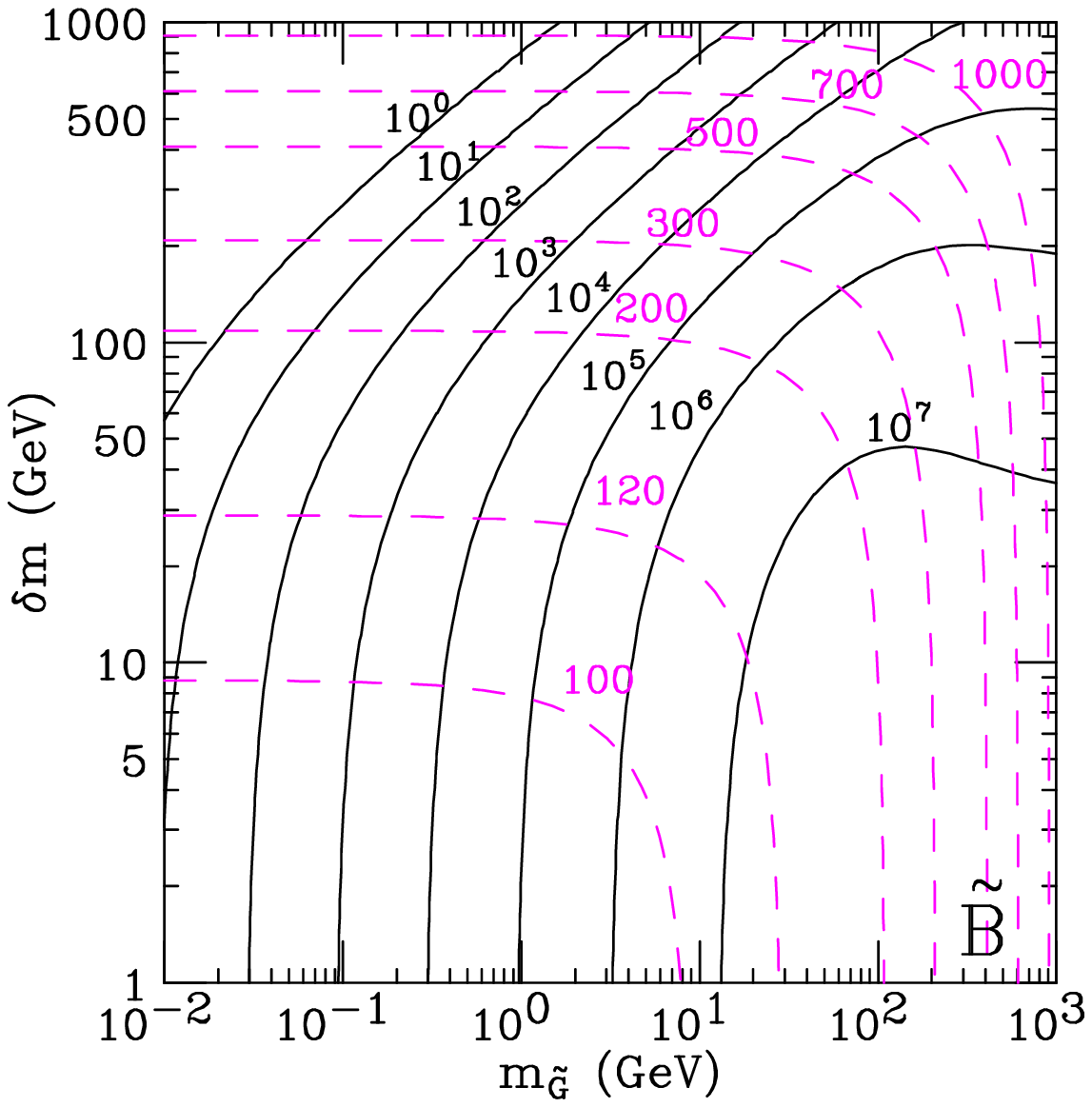}}
\caption{As in \figref{life_slep}, but for a Bino NLSP.
\label{fig:life_Bino} }
\end{figure}

The decay width to the Higgs boson is\footnote{This result disagrees
with the decay width given in Ref.~\cite{Ellis:2003dn}. After
cross-checking with the authors, they agree with our current
results. We thank Y.~Santoso and V.~Spanos for helpful
correspondence.}
\begin{eqnarray}
\Gamma (\chi \to h \gravitino) &=&
\frac{| - \bold{N}_{13} \sin \alpha + \bold{N}_{14} \cos \alpha |^2}
{96\pi M_*^2} \frac{\mchi^5}{\mgravitino^2}
F(\mchi,\mgravitino,m_h) \nonumber \\
&& \times \left[ \left( 1-\frac{\mgravitino}{\mchi} \right)^2
\left( 1 + \frac{\mgravitino}{\mchi} \right)^4
-\frac{m_h^2}{\mchi^2}  H (\mchi,\mgravitino,m_h) \right] \ ,
\label{hwidth}
\end{eqnarray}
where $h =(- H_d^0 \sin \alpha + H_u^0 \cos \alpha)/\sqrt{2}$, $F$ is
as given in \eqref{F}, and
\begin{eqnarray}
H(\mchi,\mgravitino,m_h) &=& 3 + 4\frac{\mgravitino}{\mchi}
+2 \frac{\mgravitino^2}{\mchi^2}+4 \frac{\mgravitino^3}{\mchi^3}
+3\frac{\mgravitino^4}{\mchi^4}
+ \frac{m_h^4}{\mchi^4} \nonumber \\
&& - \frac{m_h^2}{\mchi^2}
\left(3 +2 \frac{\mgravitino}{\mchi}
+3\frac{\mgravitino^2}{\mchi^2} \right) \ .
\end{eqnarray}
For the case of a Bino-like neutralino, the neutralino's mass and
lifetime are given in the $(m_{\gravitino}, \delta m)$ plane in
\figref{life_Bino}.

Given these two-body decay widths, the resulting values for the
energy release parameters are
\begin{eqnarray}
B_{\text{EM}}^{\chi} &\simeq& 1 \\
\epsilon_{\text{EM}}^{\chi} &=& 
\frac{m_{\chi}^2 - m_{\gravitino}^2}{2 m_{\chi}} \\
B_{\text{had}}^{\chi} &\simeq& 
\frac{\Gamma(\chi \to Z \gravitino) B_{\text{had}}^Z
+ \Gamma(\chi \to h \gravitino) B_{\text{had}}^h
+ \Gamma(\chi \to q \bar{q} \gravitino) }
{\Gamma(\chi \to \gamma \gravitino) + 
\Gamma(\chi \to Z \gravitino) + 
\Gamma(\chi \to h \gravitino)} \\
\epsilon_{\text{had}}^{\chi} &\approx& 
\frac{m_{\chi}^2 - m_{\gravitino}^2 + m_{Z,h}^2}{2 m_{\chi}} \ ,
\end{eqnarray}
where $B_{\text{had}}^h \approx 0.9$.  For the three-body decay, we
take $\Gamma ( \chi \to q \bar{q} \gravitino) \sim 10^{-3}$, which
provides a lower bound on $B_{\text{had}}^{\chi}$ when the two-body
decays become kinematically suppressed.  In addition, for $m_{\chi} -
m_{\gravitino}< m_Z$, $\epsilon_{\text{had}}^{\chi}$ is estimated to
be $\frac{2}{3} (m_{\chi} - m_{\gravitino})$ in our analyses.

We have neglected decays to the heavy Higgs bosons.  When
kinematically allowed, they will, of course, modify the branching
fraction and energy release formulae above.  The decay width to the
heavy CP-even Higgs boson $H$ is given by replacing $m_h \to m_H$ and
$- \bold{N}_{13} \sin \alpha + \bold{N}_{14} \cos \alpha \to
\bold{N}_{13} \cos \alpha + \bold{N}_{14} \sin \alpha$ in
\eqref{hwidth}.  The decay width to the CP-odd Higgs boson $A$ is
given by replacing $m_h \to m_A$, $- \bold{N}_{13} \sin \alpha +
\bold{N}_{14} \cos \alpha \to \bold{N}_{13} \sin \beta + \bold{N}_{14}
\cos \beta$ and, in the last line of \eqref{hwidth}, $m_{\chi}\to
-m_{\chi}$, where the last transformation is required by the CP-odd
nature of the $A$ boson.

\section{Thermal Relic Densities}
\label{sec:thermal}

To determine the normalized NLSP number density $Y_{\text{NLSP}}$ of
\eqref{xi} and also the resulting contribution of gravitinos to the
current dark matter energy density, we assume that the NLSP freezes
out with its thermal relic density.  The superWIMP has no effect on
the early thermal history of the Universe.  The NLSP therefore freezes
out as usual, with relic density given approximately
by~\cite{Bernstein:th,Scherrer:zt}
\begin{eqnarray}
\Omega_{\text{NLSP}}^{\text{th}} h^2 \approx 
\frac{1.1\times 10^9 \, x_F \, \gev^{-1}}
{\sqrt{g_*} \, \mplanck \, c_J \langle\sigma v\rangle} 
\approx 0.2\Biggl[\frac{15}{\sqrt{g_*}}\Biggr]
\Biggl[\frac{x_F}{30}\Biggr]
\Biggl[\frac{10^{19}~\gev}{\mplanck}\Biggr]
\Biggl[\frac{10^{-9}~\gev^{-2}}{c_J\langle\sigma v\rangle}\Biggr]\, ,
\label{relic}
\end{eqnarray}
where $x_F=m_{\text{NLSP}}/T_F$ is the NLSP mass divided by the freeze
out temperature $T_F$, $g_*$ is the effective number of massless
degrees of freedom at freeze out, and $\langle\sigma v\rangle$ is the
thermally-averaged NLSP annihilation cross section, and $c_J$ is 1 for
$S$-wave annihilation, 1/2 for $P$-wave annihilation.  The energy
release parameter $Y_{\text{NLSP}}$ is derived from this through
\begin{equation}
Y_{\text{NLSP}} = \frac{\Omega_{\text{NLSP}}^{\text{th}} \rho_c}
{m_{\text{NLSP}} n_{\gamma}^{\text{BG}}} 
\simeq 1.3 \times 10^{-11} 
\left[ \frac{\tev}{m_{\text{NLSP}}} \right]
\Omega_{\text{NLSP}}^{\text{th}} \ ,
\end{equation}
and the gravitino relic density is given by
\begin{equation}
\Omega_{\gravitino} h^2 = \frac{m_{\gravitino}}
{m_{\text{NLSP}}}\Omega_{\text{NLSP}}^{\text{th}} h^2 \ .
\end{equation}

For the case of slepton NLSPs, the dominant annihilation channels are
typically $\tilde{l} \tilde{l}^* \to \gamma \gamma, \gamma Z, ZZ$
through slepton exchange and $\tilde{l} \tilde{l} \to l l$ through
Bino exchange.  For right-handed sleptons, the thermally-averaged
cross section near threshold may be approximated
as~\cite{Asaka:2000zh}
\begin{eqnarray}
\langle \sigma v \rangle _{\tilde{l}_R} 
&\approx& \frac{4\pi \alpha^2}{m_{\tilde{l}_R}^2}+
 \frac{16\pi\alpha^2m_{\tilde{B}}^2} 
{\cos^4 \theta_W (m_{\tilde{l}_R}^2 + m_{\tilde{B}}^2 )^2}
= 5.0\times 10^{-10} \,
C\left[\frac{\tev}{m_{\tilde{l}_R}}\right]^2 \gev^{-2} \ ,
\end{eqnarray}
where $m_{\tilde{B}}$, $m_{\tilde{l}_R}$ are the Bino and slepton
masses, respectively, and $C$ is an ${\cal O}(1)$ model-dependent
constant.  Here we have not included co-annihilation processes, which
might be important if sleptons and, say, neutralinos are nearly mass
degenerate.  Using \eqref{relic} and setting $C=1$, the slepton
thermal relic abundance is
\begin{equation}
\Omega_{\tilde{l}_R}^{\text{th}} h^2 \approx 0.2 \,
\left[\frac{m_{\tilde{l}_R}}{\tev}\right]^2 \ .
\label{omegal}
\end{equation} 

A similar analysis for the sneutrino NLSP case yields~\cite{Fujii:2003nr}
\begin{equation}
\Omega_{\tilde{\nu}}^{\text{th}} h^2 \approx 0.06 \,
\left[\frac{m_{\tilde{\nu}}}{\tev}\right]^2 \ . 
\label{omeganu}
\end{equation} 
The thermal relic density of the sneutrino is typically smaller than
that of right-handed sleptons because sneutrino annihilation is
relatively efficient, taking place through weak SU(2) couplings,
whereas the right-handed sleptons annihilate only through hypercharge
couplings.

For the neutralino NLSP case, the thermal relic density is very
model-dependent.  The annihilation cross section varies widely
depending on the gaugino-Higgsino composition of the neutralino and
the presence or absence of co-annihilation effects, and so depends on
a large number of unknown supersymmetry parameters.  Rather than
constraining these parameters by working in a particular model, we
adopt a simple scaling behavior based on some well-known results.  In
particular, we assume that the annihilation cross section scales as
$m_{\chi}^{-2}$.  To fix the constant of proportionality, we recall
that in the ``bulk'' region of minimal supergravity, where the
neutralino is Bino-like and there is no significant co-annihilation,
the desired relic density is achieved for $m_{\tilde{B}} \approx
100~\gev$.  In the focus point (FP) region of minimal
supergravity~\cite{Feng:1999hg,Feng:1999mn,Feng:1999zg}, where the
neutralino is a Bino-Higgsino mixture~\cite{Mizuta:1992qp}, the
neutralino mass may be much larger~\cite{Feng:2000gh,Feng:2000zu}.  If
there are co-annihilation
effects~\cite{Binetruy:1983jf,Griest:1990kh}, the neutralino mass may
also be much higher~\cite{Ellis:1999mm,Arnowitt:2001yh,Nihei:2002sc}.
To study the effect of having a heavier neutralino, we consider the
mass $m_{\tilde{B}} \approx 200~\gev$ as an example of these other
possibilities.  We therefore consider the range
\begin{equation}
\text{``bulk'':} \quad \Omega_{\chi}^{\text{th}} h^2 \approx 
0.1 \, \left[\frac{m_{\chi}}{100~\gev}\right]^2 
\label{omegachi1}
\end{equation} 
to
\begin{equation}
\text{``focus point/co-annihilation'':} \quad 
\Omega_{\chi}^{\text{th}} h^2 \approx 
0.1 \, \left[\frac{m_{\chi}}{200~\gev}\right]^2 \ .
\label{omegachi2}
\end{equation} 

Note that for similar NLSP masses, the thermal relic density is much
higher in the neutralino case than in the slepton case.  This is as
expected, because the neutralino annihilation is dominantly $P$-wave
because of the Majorana-ness of neutralinos, while slepton
annihilation takes place in the $S$-wave.  In fact, for similar
masses, one expects the slepton relic density to be suppressed
relative to the neutralino relic density by a factor of roughly $v^2
\sim 3/x_F \sim 1/10$.  Given the approximations used, this is in
reasonable quantitative agreement with the estimates of
Eqs.~(\ref{omegal})--(\ref{omegachi2}).

\section{Constraints}
\label{sec:constraints}

\subsection{Dark Matter Density}

An unambiguous and simple constraint on these scenarios is that the
resulting gravitino energy density should not be greater than the
observed dark matter density.  This constraint may be avoided if there
is significant entropy production between NLSP freeze out and now.
However, assuming such new physics is counter to our goal of
evaluating the gravitino LSP possibility in the simplest possible
cosmology, and so we require
\begin{equation}
\Omega_{\gravitino} h^2 = \frac{m_{\gravitino}}
{m_{\text{NLSP}}}\Omega_{\text{NLSP}}^{\text{th}} h^2 < 0.11 \ .
\end{equation}

As evident from \eqsref{sfermionlifetime}{neutralinolifetime} and
\figsref{life_slep}{life_Bino}, the typical decay times are $t \sim
10^4 - 10^8~\s$.  This is the natural decay time of a particle with
weak-scale mass that decays through gravitational interactions.  There
are therefore additional constraints, most importantly from bounds on
EM energy release from cosmic microwave background (CMB) $\mu$
distortions and from bounds on both EM and hadronic energy release
from BBN light element abundances.

\subsection{Cosmic Microwave Background}

The CMB constraint is fairly straightforward to understand.  The CMB
photon energy distribution at times $t \alt 10^8~\s$ may be
parameterized as
\begin{equation}
f_{\gamma}(E) = \frac{1}{e^{E/(kT) + \mu} - 1} \ ,
\end{equation}
with chemical potential $\mu$.  For early decays, EM cascades are
completely thermalized through energy-changing processes $\gamma e^-
\to \gamma e^-$ and number-changing interactions, such as $e X \to e X
\gamma$, where $X$ is an ion, and double Compton scattering $\gamma
e^- \to \gamma \gamma e^-$.  The resulting distribution is therefore
Planckian, with $\mu = 0$.  For decay times in the window of interest,
however, the number-changing processes may be inefficient.  In this
case, the spectrum cannot relax to a distribution determined by only
one parameter, the temperature $T$.  It therefore relaxes to
statistical but not thermodynamic equilibrium, resulting in a
Bose-Einstein distribution function with $\mu \ne 0$.

The value of the chemical potential $\mu$ may be approximated for
small energy releases by analytic expressions given in
Ref.~\cite{Hu:gc}.  These have been updated with current cosmological
parameters in Ref.~\cite{Feng:2003uy}.  We will apply the current
constraint~\cite{Fixsen:1996nj,Hagiwara:fs}
\begin{equation}
|\mu| < 9 \times 10^{-5} \ .
\label{mu}
\end{equation}

\subsection{Big Bang Nucleosynthesis}
\label{sec:BBN}

The BBN constraints are more complicated and more ambiguous.
Constraints on EM energy release have been studied
in~\cite{Ellis:1984er,Ellis:1990nb,Kawasaki:1994sc,Holtmann:1998gd,%
Kawasaki:2000qr,Asaka:1998ju}.  Most recently, EM constraints (but not
hadronic constraints) have been considered in
Ref.~\cite{Cyburt:2002uv} and these were used in the previous analyses
of Refs.~\cite{Feng:2003xh,Feng:2003uy}.  Here we include contours
corresponding to the most stringent constraint from that analysis, the
deuterium bound
\begin{equation}
1.3 \times 10^{-5} \ \ < \ \ \text{D/H} \ \ < \ \ 5.3\times 10^{-5} \ ,
\label{DEllis}
\end{equation}
to facilitate comparison with previous results.

More recently, both EM and hadronic energy releases have been bounded
in the analysis of Ref.~\cite{Kawasaki:2004yh}.  Of the constraints
imposed there, the most relevant for us are the 2$\sigma$ bounds
\begin{eqnarray}
2.4 \times 10^{-5} \ \ < &\text{D/H}& < \ \ 3.2\times 10^{-5} \label{D} \\
&\text{$^3$He/D}& < \ \ 1.13 \label{3He} \\
&\text{$^6$Li/H}& < \ \ 6.1 \times 10^{-11} \label{6Li} \\
0.228 \ \ < &Y_p& < \ \ 0.248 \label{4He} \  . 
\end{eqnarray}
The statistical and systematic errors have been combined in quadrature
for the $^4$He ($Y_p$) constraint, and the $^6$Li/H result is obtained
by combining the 95\% confidence level (CL) constraints $^6\text{Li}/^7\text{Li} = 0.05
\pm 0.02$ and $^7\text{Li}/\text{H} = 2.2_{-1.6}^{+6.5} \times
10^{-10}$~\cite{Kohri}.

As is evident, the later analysis is much less conservative.  First,
it assumes a significantly more stringent D bound.  Measurements of
primordial D have long been considered by many to be the most reliable
baryometers.  There is also now impressive concordance between the
baryon number determinations from D and CMB measurements, which
further supports the narrow range of D/H given in \eqref{D}.  At the
same time, existing discrepancies between standard BBN predictions and
observations in other elements may indicate that caution is still
needed in interpreting the D bound.  In particular, if these
discrepancies are indications of new physics, the required new physics
is also likely to distort the D abundance, since the D binding energy
is so small.

The analysis of Ref.~\cite{Kawasaki:2004yh} also includes stringent
constraints from $^3$He/D and $^6$Li/H.  If taken at face value, these
additional bounds in fact provide some of the most stringent bounds on
the gravitino LSP scenario. The $^3$He/D bound is the strongest
constraint on EM energy and provides the strongest constraint for NLSP
decay times $\tau \agt 10^7~\s$, while the $^6$Li/H constraint on
hadronic energy release provides the strongest constraint on earlier
decay times.  At the same time, it is important to bear in mind that
these constraints are on less sure footing than the D constraints.
For $^3$He, $^3$He/H suffers from uncertainties in chemical/stellar
evolution~\cite{Vangioni-Flam:2002sa}.  Although $^3$He/D has been
proposed as an alternative tool to constrain new
physics~\cite{Sigl:1995kk,Holtmann:1998gd}, present evaluations exist
only in the Sun~\cite{Scully:1995tp}, and the determination of
primordial $^3$He/D requires a rather involved extrapolation of these
results.  $^6$Li has also been proposed as a promising probe of new
physics.  However, after WMAP, there is a clear discrepancy between
standard BBN predictions and the observations of
$^7$Li~\cite{Thorburn,Bonafacio,Ryan:1999vr}, with consistency
possible only if systematic uncertainties have been
underestimated~\cite{Cyburt:2003ae}.  This calls the status of $^6$Li
into question, as direct observations of $^6$Li/H are difficult, and
so the upper limit on $^6$Li/H is usually derived from bounds on
$^6$Li/$^7$Li.  In fact, the current status of $^6$Li and $^7$Li may
also be taken as evidence for new particle
physics~\cite{Jedamzik:1999di,Jedamzik:2004er}.

In light of all of these comments, we present constraint contours from
all of these data, including both D constraints, to show the (strong)
effect of varying BBN assumptions.  We consider regions of parameter
space that violate the conservative constraint of \eqref{DEllis} to be
excluded, but we consider regions that violate only
Eqs.~(\ref{D})--(\ref{6Li}) to be at most disfavored, but not
necessarily excluded, given the significant ambiguities noted above.

There are subtleties in importing the constraints of
Eqs.~(\ref{D})--(\ref{6Li}) to the present analysis.  When including
both EM and hadronic energy release, there is the possibility of
cancellations.  In addition, although the EM constraint depends
essentially only on $\xi_{\text{EM}}$ of \eqref{xi}, the hadronic
constraint may depend, in principle, $\epsilon_{\text{had}}$ and
$B_{\text{had}} Y_{\text{NLSP}}$ separately, and results are presented
only for a few values of $\epsilon_{\text{had}}$.  In practice,
however, the cancellations occur only in rather special cases for
particular energy release time.  In addition, in supergravity with a
gravitino LSP, the NLSP lifetime is usually larger than 150 sec, and
so the hadronic constraint depends to a good approximation on
$\xi_{\text{had}} \equiv \epsilon_{\text{had}} B_{\text{had}}
Y_{\text{NLSP}}$ only.\footnote{The constraint on $\xi_{\text{had}}$
varies only by a factor of 2 for $\epsilon_{\text{had}}$ between 100
GeV and 1 TeV~\cite{Kohri}.}  We therefore impose constraints on
$\xi_{\text{had}}$ and impose the constraints on EM and hadronic
energy release separately, ignoring the possibility of cancellations.

\section{Results}
\label{sec:results}

We have now determined the energy release parameters $B_{\text{i}}$
and $\epsilon_{\text{i}}$ in \secref{late} and $Y_{\text{NLSP}}$ in
\secref{thermal}.  We may now compare these to the constraints of
\secref{constraints} to determine what combinations of gravitino mass
and NLSP mass are excluded, disfavored, and allowed for various NLSP
possibilities. We will present results in the $(\mgravitino, \delta m
)$ plane, where $\delta m \equiv m_{\text{NLSP}} - \mgravitino - m_Z$.
We consider only $\delta m > 0$, so three-body decays are therefore
always kinematically possible.  Of course, they are highly suppressed
for small $\delta m$.

\subsection{Slepton NLSP}

We begin with the stau NLSP scenario. We assume the stau is
right-handed.  Neutralino and chargino parameters enter in the
three-body decay widths.  We take $\mu = M_2 = 2 M_1 = 4 m_{\stau_R}$
and $\tan\beta = 10$.

\begin{figure}
\resizebox{6.5 in}{!}{
\includegraphics{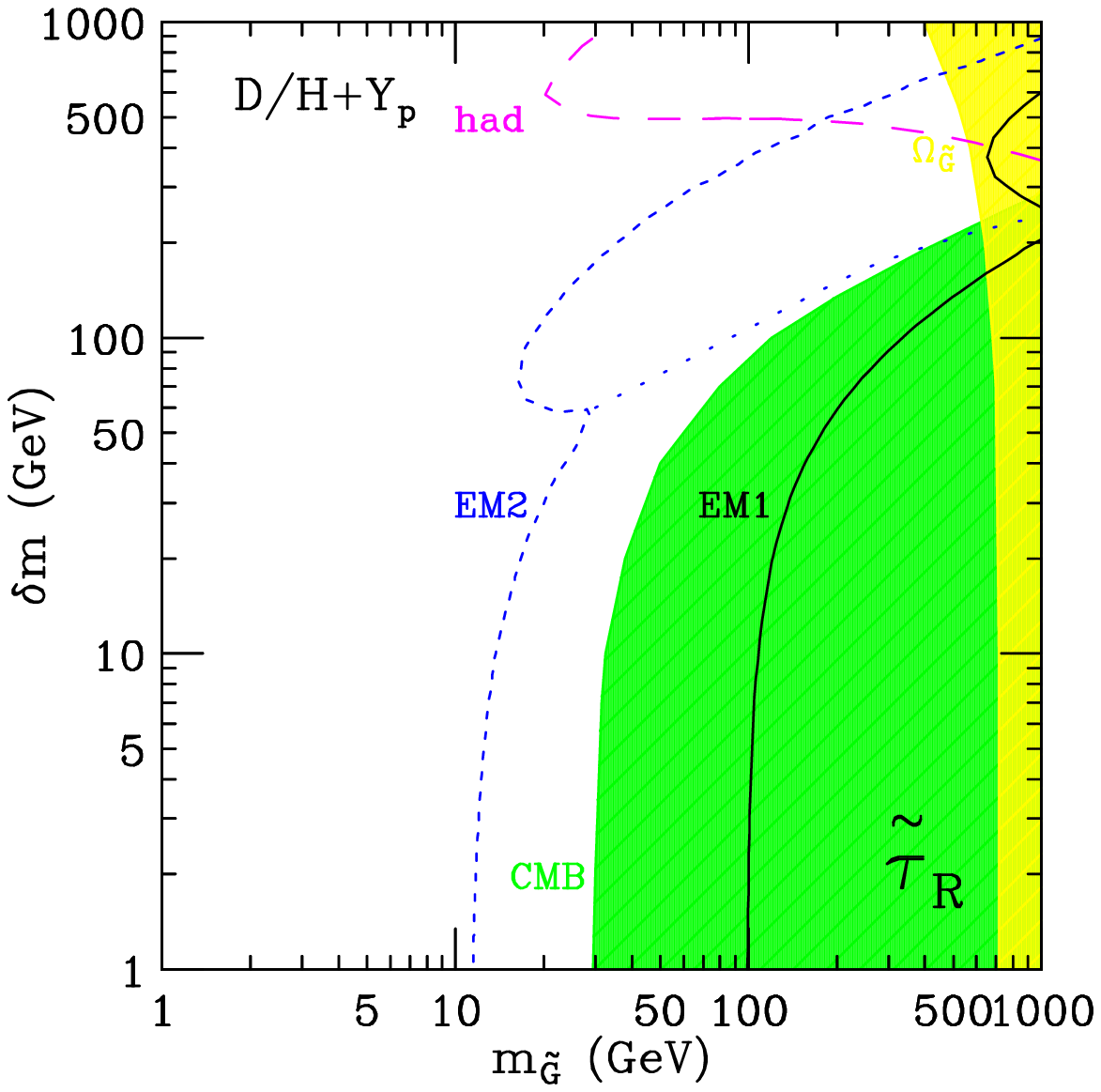}
\includegraphics{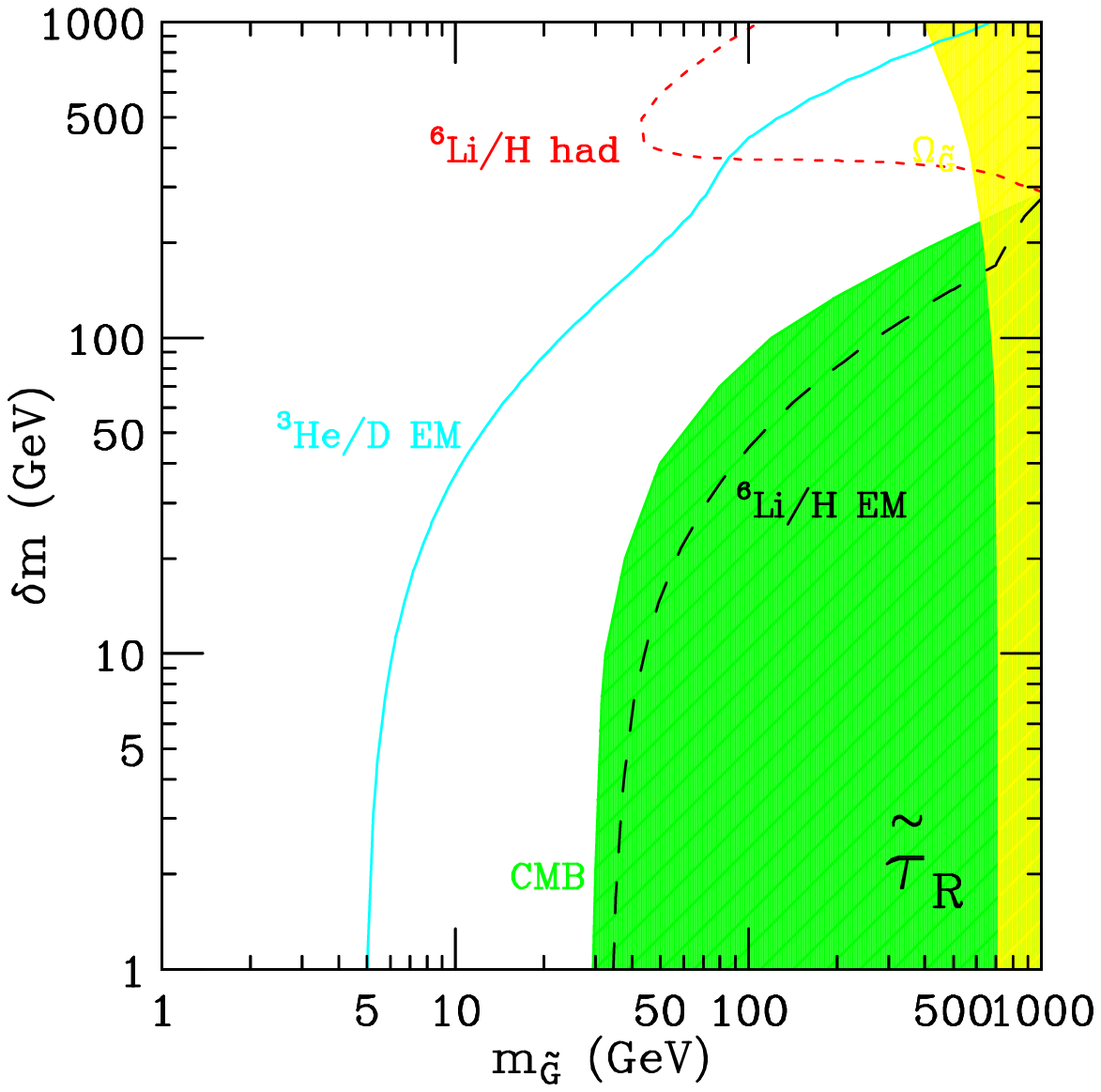}}
\caption{Excluded and allowed regions of the $(m_{\gravitino}, \delta
m \equiv m_{\text{NLSP}} - m_{\gravitino} - m_Z)$ parameter space in
the gravitino LSP scenario, assuming a $\tilde{\tau}_R$ NLSP that
freezes out with thermal relic density given by \eqref{omegal}. The
light (yellow) shaded region is excluded by the overclosure constraint
$\Omega_{\gravitino} h^2 < 0.11$, and the medium (green) shaded region
is excluded by the absence of CMB $\mu$ distortions.  BBN is sensitive
to the regions to the right of the labeled contours.  Left: Regions
probed by D and $^4$He, assuming the conservative result of
\eqref{DEllis} (EM1), and the more stringent constraints of
\eqsref{D}{4He} (EM2 and had).  The dotted line denotes the region
where cancellation between D destruction and creation via late time EM
injection is possible~\cite{Dimopoulos:1988ue}, while ${}^7$Li is
reduced to the observed value by the late NLSP
decays~\cite{Cyburt:2002uv}.  Right: Regions probed by $^3$He/D (EM),
$^6$Li/H (had)~\cite{Kawasaki:2004yh}, and $^6$Li/H
(EM)~\cite{Kawasaki:2004yh, Cyburt:2002uv}.
\label{fig:stau} }
\end{figure}

The results are presented in \figref{stau}.  To understand these
results, it may be helpful to refer to the mass and lifetime contours
of \figref{life_slep}.  Note that, given the definition of $\delta m$,
$m_{\stau_R} > m_Z$ in the entire plane.  The current limit on a
metastable stau from LEP is $m_{\stau_R} > 99~\gev$~\cite{leptrack},
and so excludes a small portion of the lower lefthand corner.

The shaded regions are excluded.  Given the scaling
$\Omega_{\NLSP}^{\text{th}} \propto m_{\NLSP}^2$ for the thermal relic
density, the dark matter density implies an upper bound on the product
$m_{\NLSP} \mgravitino$, excluding $\stau$ and gravitino masses $\sim
1 ~\tev$.  The constraint is relatively mild, because staus annihilate
efficiently through $S$-wave processes.  The other shaded region is
excluded by the absence of CMB $\mu$ distortions.  This provides a
more stringent constraint than $\Omega_{\gravitino}$ for decay times
$\tau \agt 10^7~\s$, when the decay products are produced too late to
be thermalized.

The BBN sensitivity contours divided into those from D and $^4$He,
which are probably the most reliable (left panel), and those from
$^3$He and $^6$Li, which are on less sure footing, given the
discussion of \secref{BBN}.  For the D and $^4$He results, we present
results given the conservative bound of \eqref{DEllis} on EM cascades
(EM1), the more aggressive bound of \eqsref{D}{4He} on EM energy
(EM2), and the bound of \eqsref{D}{4He} on hadronic cascades (had).
The EM1 contour lies completely in the CMB-excluded region.  Although
BBN constraints are often assumed to be the leading constraint on late
decays, we see that the CMB spectrum is now know to be Planckian to
such high precision that the CMB constraint is competitive with the
leading BBN constraints.  At the same time, we see that the strength
of the EM constraints in constraining gravitino LSP parameter space
depends sensitively on how one interprets the BBN data.  Adopting the
more stringent EM2 contour, we find that bounds on $\mgravitino$ are
improved by about an order of magnitude.

Our analysis includes hadronic bounds on the gravitino LSP scenario
for the first time.  {}From \figref{stau}, we see that the hadronic
constraint is the leading constraint for relatively early decays.
Recall that sleptons produce hadronic energy only in three-body
decays, and so the hadronic energy release is suppressed by factors of
$\sim 10^{-3}$ relative to EM energy.  Nevertheless, hadronic decay
products are so lethal to light elements that the hadronic constraints
are the most stringent constraint in parts of parameter space.  Note
also that the part of parameter space in which hadronic constraints
are most important is where $m_{\stau_R}$ and $\mgravitino$ are both
in the hundreds of GeV, the most natural region for weak-scale
supergravity.  We conclude that hadronic constraints and three-body
decays must be taken into account to establish the viability of any
gravitino LSP scenario.

In the right panel of \figref{stau}, we include the sensitivity
contours of $^3$He and $^6$Li.  We see that, taken literally, the
constraint on $^3$He provides the most stringent constraint on late
decays (through its limits on EM energy) and $^6$Li provides the
leading constraint on early decays (through its limits on hadronic
energy).  Of course, given the ambiguities discussed in \secref{BBN},
we do not consider these contours to be exclusion contours.  These
sensitivity contours are of interest, however, as, if colliders
measure the superpartner parameters to be in the regions to the right
of these contours (as we will discuss in \secref{collider}), these
measurements will have important implications for BBN.

Last, we discuss the dark matter implications of these results.  At
the boundary of the region excluded by $\Omega_{\gravitino}$, the
light (yellow) shaded region, gravitino superWIMPs account for all of
dark matter.  Taking the EM1 and hadronic constraints from D and
$^4$He, we see that this possibility is indeed viable.  If, however,
the various BBN anomalies are resolved and the EM2 and $^3$He and
$^6$Li contours may be considered as exclusion contours, the
possibility that superWIMP gravitinos form all of dark matter may be
excluded.  However, given the current status of BBN, we find such
conclusions premature.

Finally, we should remind the reader that we have assumed a particular
thermal relic density and particular neutralino mass parameters, which
enter the three-body branching ratios.  All of the contours above will
shift if there are significant deviations in these assumptions.  A
complete analysis of all of these variations is, however, beyond the
scope of this work.

\subsection{Sneutrino NLSP}

As discussed in \secref{sneutrino}, if the NLSP is a sneutrino,
two-body decays are essentially invisible, and so inclusion of
three-body decays is essential to determine the viable parameter
space.  Because the hadronic constraints are some much stronger than
the EM constraints, we may focus on them only.  Again, neutralino and
chargino parameters enter in the three-body decay widths, and we
assume $\mu = M_2 = 2 M_1 = 4 m_{\tilde{\nu}}$ and $\tan\beta = 10$.

\begin{figure}
\resizebox{3.25 in}{!}{
\includegraphics{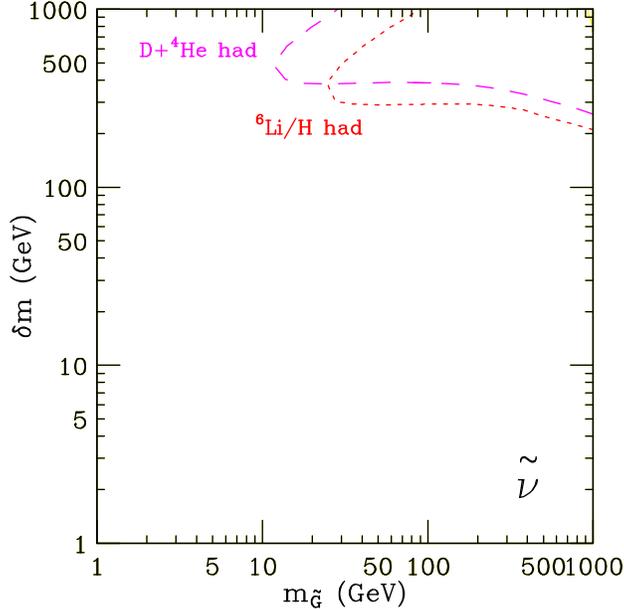}}
\caption{As in \figref{stau}, but assuming a sneutrino NLSP that
  freezes out with thermal relic density given by \eqref{omeganu}.
\label{fig:sneu} }
\end{figure}

The results are presented in \figref{sneu}.  Sneutrinos annihilate
through $S$-wave processes even more efficiently than sleptons, as can
be seen by comparing \eqsref{omegal}{omeganu}.  The dark matter
density bound is therefore weaker, and is in fact pushed to the right;
it does not appear in the plotted plane.  The CMB constraint,
previously so stringent, is also absent, of course, as it constrains
EM energy, and the BBN constraints on EM energy are also absent.

The remaining constraints are therefore only the hadronic BBN
constraints.  These are stringent for early decays, that is, large
$\delta m$.  The more reliable D and $^4$He constraints disfavor
$\delta m \agt 300~\gev$, while $^6$Li (had) is sensitive to $\delta m
\agt 200~\gev$.  It is rather remarkable that the sneutrino NLSP case
is so tightly constrained, given its invisible dominant decay mode.
At the same time, the scenario is perfectly viable for natural
weak-scale supergravity parameters.  Note that gravitino superWIMP
dark matter is also viable for $\mgravitino \sim 1~\tev$ and $\delta m
\alt 300 ~\gev$.

\subsection{Bino NLSP}

\begin{figure}
\resizebox{6.5 in}{!}{
\includegraphics{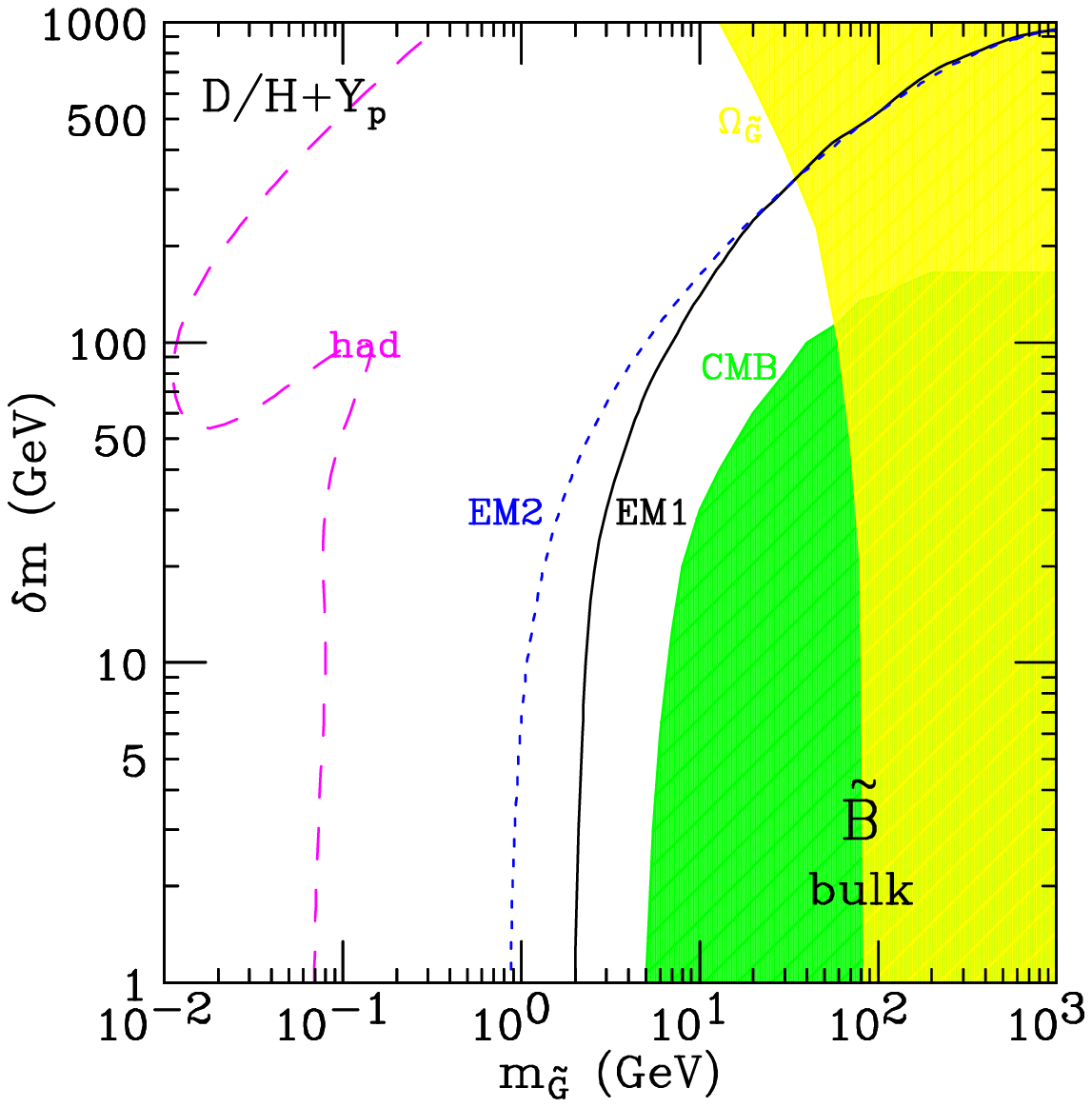}
\includegraphics{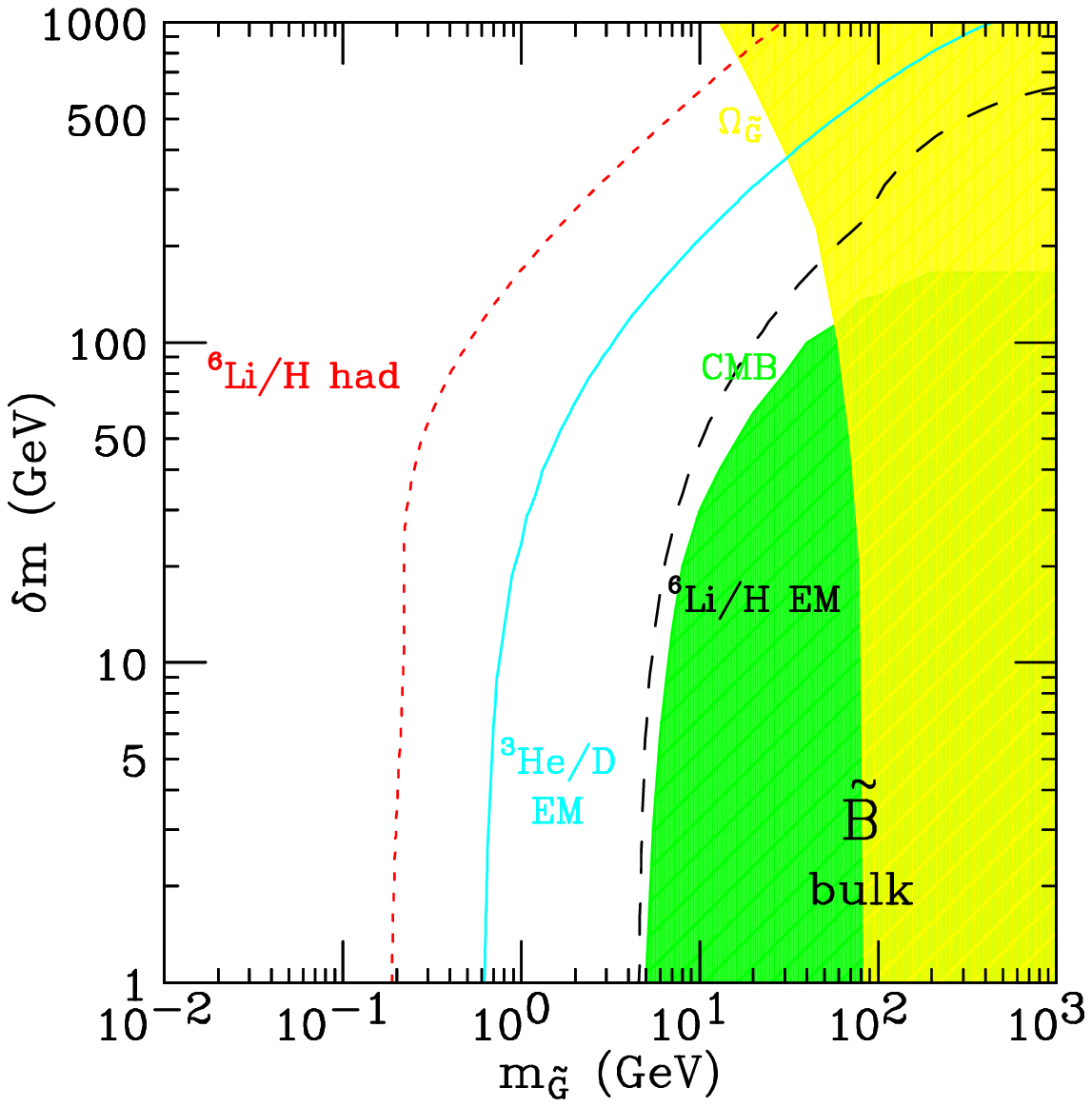}}
\caption{As in \figref{stau}, but assuming a Bino NLSP that freezes
out with the ``bulk'' thermal relic density given by
\eqref{omegachi1}. 
\label{fig:Bino_bulk} }
\end{figure}

\begin{figure}
\resizebox{6.5 in}{!}{
\includegraphics{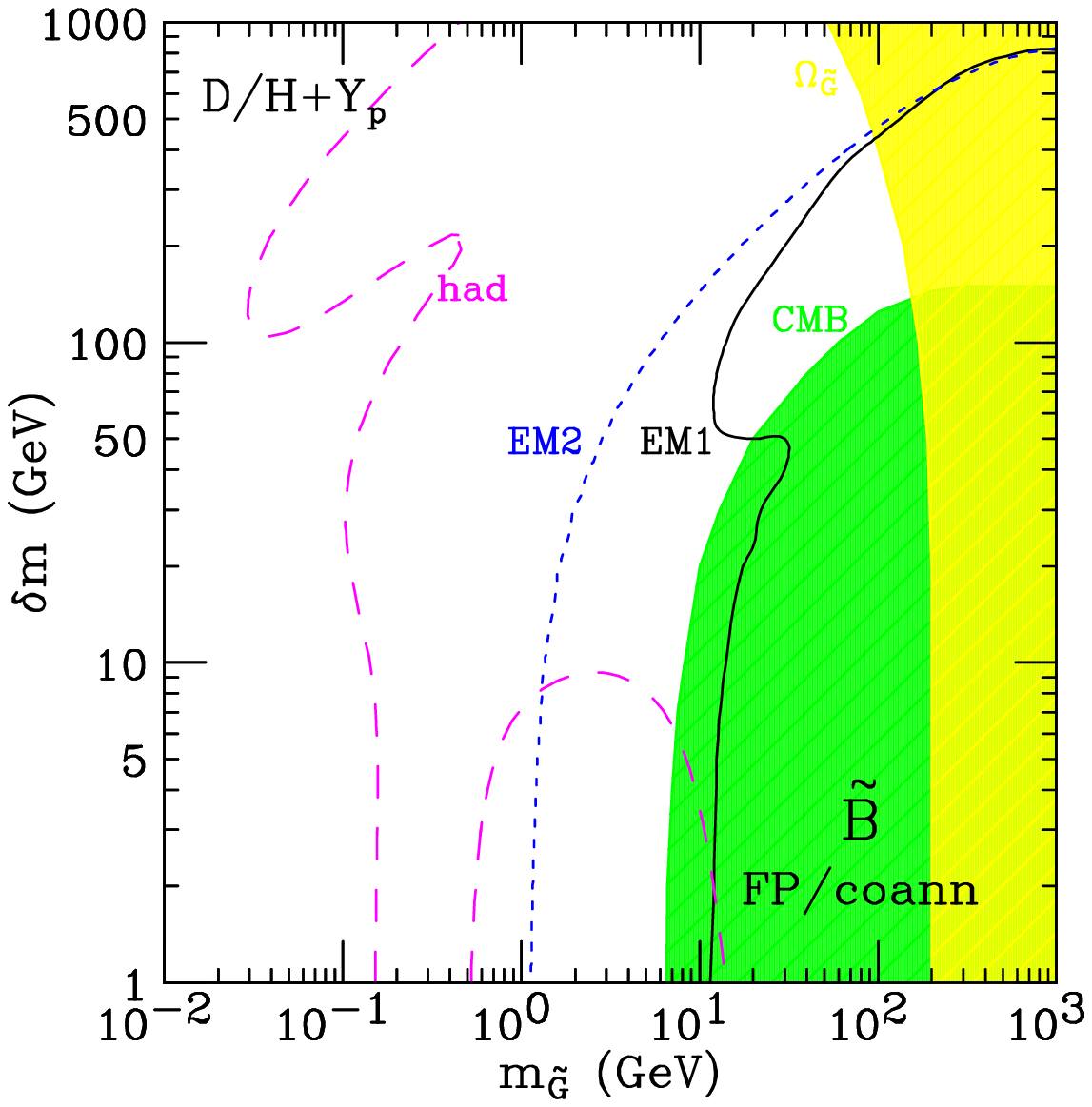}
\includegraphics{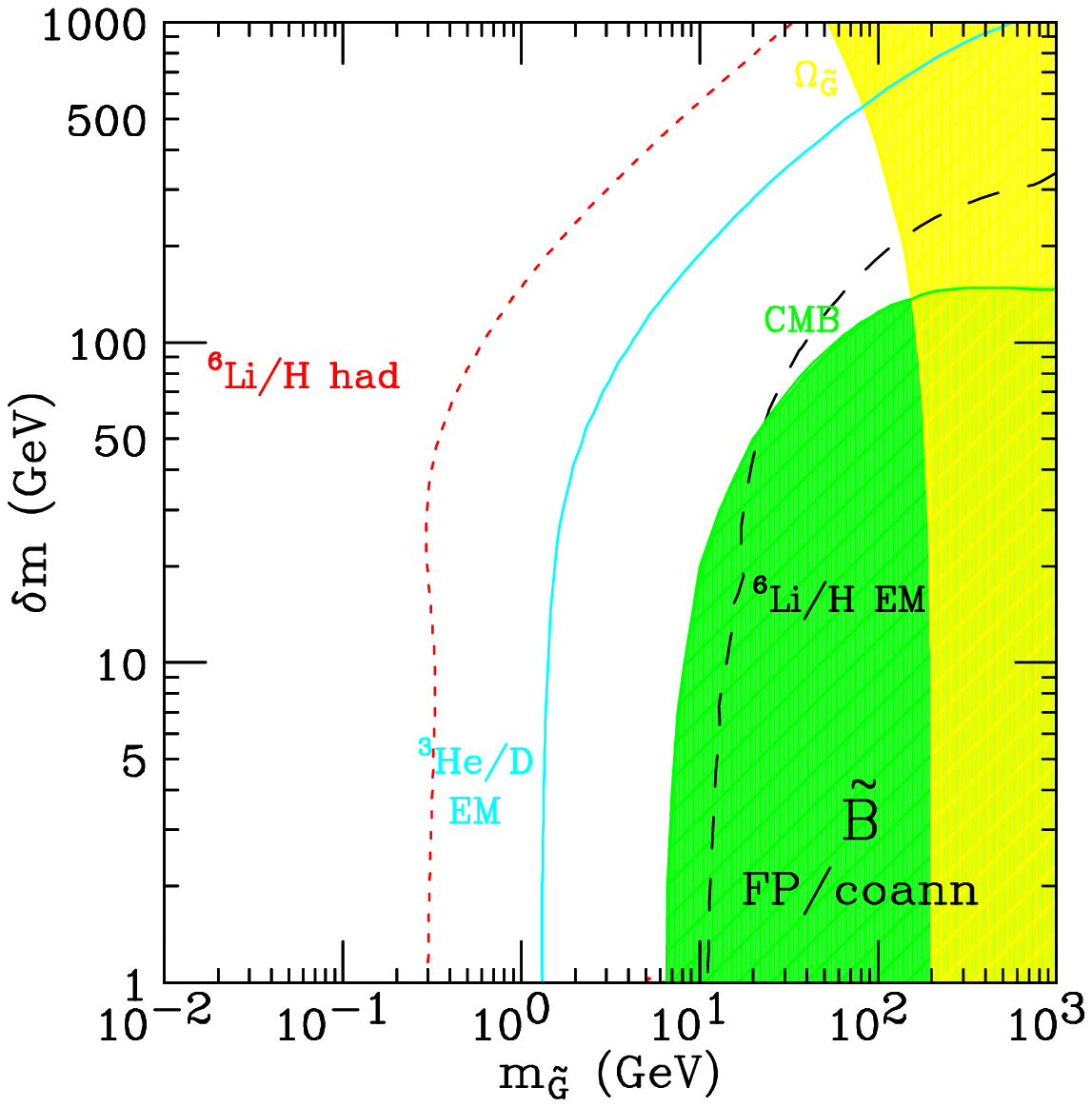}}
\caption{As in \figref{stau}, but assuming a Bino NLSP that freezes
out with the ``focus point/co-annihilation'' thermal relic density
given by \eqref{omegachi2}. In the left panel, the region between the
``had'' lines is disfavored.
\label{fig:Bino_fp} }
\end{figure}

Finally, we turn to the case of the Bino NLSP.  The results are
presented in \figref{Bino_bulk} for the case where the Bino thermal
relic density is as in the bulk region of minimal supergravity
(\eqref{omegachi1}), and in \figref{Bino_fp} for the case where the
Bino relic density is degraded by a factor of 4 (\eqref{omegachi2}),
as might be the case if there there are additional effects as may be
found in the focus point or co-annihilation regions of minimal
supergravity.  Note that the $\mgravitino$ lower limit has been
extended to much lower masses than in the slepton and sneutrino
figures.  The mass and lifetime contours of \figref{life_Bino} may be
helpful in understanding these results.

The CMB and EM BBN bounds are roughly similar to those in the slepton
NLSP case.  However, for the other bounds, there are important
changes.  Because neutralino annihilation is $P$-wave suppressed, the
dark matter density limit is much more stringent, excluding gravitino
masses above 100 GeV and 200 GeV in the ``bulk'' and ``FP/coann''
cases, respectively.  

Even more striking, the hadronic BBN bounds become much more
stringent.  This is expected --- Bino NLSP decays contribute to
hadronic energy at the two-body level through decays $\tilde{B} \to Z
\gravitino$.  The branching fractions for EM and hadronic decays are
therefore not too different, and the extreme stringency of the
hadronic constraints makes them the dominant bound.  Taking only the
relatively reliable D and $^4$He bounds, we find that decay times
$\tau \agt 10^3~\s$ are disfavored, excluding almost all gravitino
masses $\mgravitino \agt 100~\mev$.  In the ``FP/coann'' case, for
$\delta m \alt {\cal O}(10~\gev)$, the hadronic constraints exclude
$100~\mev \alt \mgravitino \alt 500~\mev$, but $\mgravitino \sim
1~\gev$ is again allowed.  In the allowed region, the decay times are
so long $\tau \agt 10^5~\s$ that the hadronic constraints become less
stringent.  However, for such long decays, the EM constraints are
stringent.  As can be seen in the figures, the EM constraints disfavor
or exclude this island of parameter space, depending on how
conservative one's interpretation of the EM BBN constraints is.

We therefore conclude that hadronic BBN constraints essentially
exclude supergravity with a gravitino LSP and a Bino NSLP when the
decay channel $\tilde{B} \to Z \gravitino$ is open.  The Bino NLSP
scenario is viable only if the Bino and gravitino masses are
degenerate enough to suppress this decay mode, if $\mgravitino$ is
below 10 MeV, a rather unnatural value of conventional supergravity,
or, possibly, if there are extremely co-annihilation effects which
suppress the thermal relic density even more than in our ``FP/coann''
example.  If the neutralino NLSP is not pure Bino, there are
additional possibilities.  For example, as noted in
Ref.~\cite{Feng:2003uy}, photino NLSPs may be viable, as they
contribute to hadronic cascades only through three-body decays.  Of
course, from the high energy viewpoint, a photino-like neutralino is
unmotivated.  More likely is the case of a Higgsino-gaugino
neutralino, for which the thermal relic density may also be greatly
suppressed, but the hadronic constraints are stronger, since $\Gamma (
\chi \to h \gravitino)$ is larger.  A detailed examination of such
focus point or co-annihilation cases would be interesting, but is
beyond the scope of this study.

\section{Implications for Collider Physics}
\label{sec:collider}

The possibility of a gravitino LSP in supergravity has rich
implications for current and future colliders.  These implications
depend crucially on whether the NLSP is a slepton, sneutrino, or
neutralino.  In all cases, however, given NLSP rest lifetimes of $10^4
- 10^8~\s$, the typical NLSP decay lengths are enormous relative to
collider detectors, and so these NLSPs are essentially stable as far
as colliders are concerned.

\subsection{Slepton NLSP}

We begin by discussing the slepton NLSP case.  This possibility is
very natural from the point of view of high-energy frameworks.  Given
simple boundary conditions at the grand unified scale, for example,
and evolving these to the weak scale, right-handed sleptons, in
particular, right-handed staus, often emerge as the lightest standard
model superpartner.  In conventional studies of these high energy
frameworks, such regions of parameter space are excluded by bounds
from searches for charged massive stable particles in sea water.  In
the gravitino LSP scenario, however, the lightest slepton is
metastable but not absolutely stable, and so these bounds do not
apply.

\begin{figure}
\resizebox{3.25 in}{!}{
\includegraphics{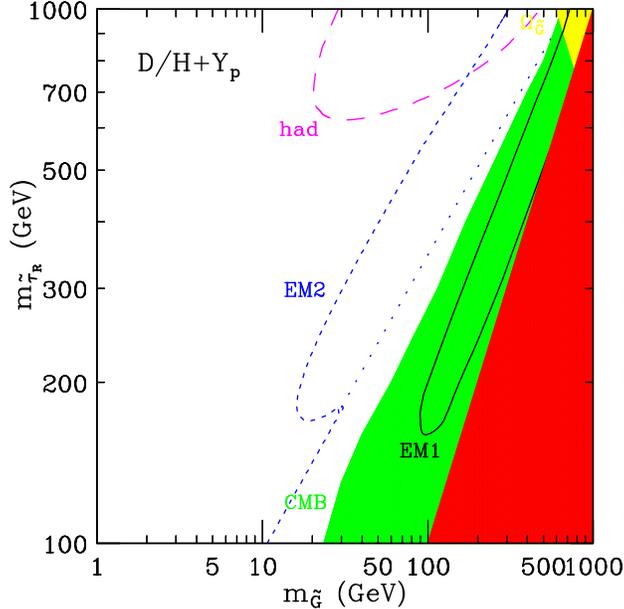}}
\caption{Excluded and allowed regions for the gravitino LSP scenario,
assuming a $\tilde{\tau}_R$ NLSP, as in the left panel of
\figref{stau}, but now in the $(m_{\gravitino}, m_{\stau_R})$ plane.
In the dark (red) shaded region the gravitino is not the LSP.  All
other shaded regions and contours are as in \figref{stau}.
\label{fig:summary_stau} }
\end{figure}

The excluded and allowed regions of parameter space are presented
again in \figref{summary_stau}, but now in the $(m_{\gravitino},
m_{\stau_R})$ plane, which is more convenient for inferring
implications for colliders.  We see that, allowing the gravitino to be
as light as 10 GeV, all weak-scale stau masses are allowed.  Staus may
therefore be within reach of the LHC and even of the first stage of a
linear collider.  For heavier gravitino masses, the allowed stau mass
range becomes more narrow.  Neglecting the aggressive EM2 bound, we
see that all dark matter may be in the form of gravitino superWIMPs if
staus have masses $m_{\stau_R} \agt 1~\tev$.  In this case, direct
stau production is beyond the range of the LHC and linear collider,
but staus may still be produced in the cascade decays of squarks and
gluinos.

At hadron colliders, sleptons can be pair-produced through the
Drell-Yan processes
\begin{eqnarray}
 &&q\bar{q}' \to W^*\to \tilde{l}_L\overline{\tilde{\nu}_L} \\
 &&q\bar{q}\to Z^* , \gamma^*\to \tilde{l}_L\overline{\tilde{l}_L}, \ 
\tilde{l}_R\overline{\tilde{l}_R},\ 
\tilde{\nu}_L\overline{\tilde{\nu}_L} \ .
\end{eqnarray}
The cross sections for such processes are determined by the slepton
masses, with very little other model dependence.\footnote{Much larger
cross sections may result from sleptons produced in cascade decays of
gluinos and squarks, but the details of these processes are highly
model-dependent.}  These Drell-Yan cross sections have been studied in
detail~\cite{Feng:1997zr,Baer:1997nh,Beenakker:1999xh}, including the
leading QCD corrections.  For the Tevatron with $\sqrt{s}=2~\tev$ and
$m_{\tilde{l}} = 100~\gev$, the cross sections for
$\tilde{l}_R\tilde{l}_R$, $\tilde{l}_L\tilde{l}_L$
($\tilde{\nu}_L\tilde{\nu}_L$), and $\tilde{l}_L\tilde{\nu}_L$ are
about 10 fb, 30 fb, and 100 fb, respectively.  These cross sections
drop quickly for heavier sleptons.  For $m_{\tilde{l}}=200~\gev$, they
are reduced by more than an order of magnitude, and sleptons with mass
above around 250 GeV will be beyond the reach of the Tevatron.

For the LHC with $\sqrt{s}=14~\tev$, the Drell-Yan cross sections are
about 10 times bigger.  Sleptons may also be produced via weak boson
fusion~\cite{Choudhury:2003hq},
\begin{equation}
qq' \to   qq^{\prime}VV \to qq' \tilde{l} \bar{\tilde{l}} \ .
\end{equation}
This cross section decreases much more slowly with increasing slepton
mass than the Drell-Yan cross section, and the weak boson fusion cross
section dominates for $m_{\tilde{l}} \agt 200 - 300~\gev$.  At the
LHC, hundreds to thousands of sleptons could be produced.

Metastable sleptons will appear as charged tracks in the tracking
chamber with little calorimeter activity.  Eventually they will hit
the muon chambers and so look muon-like.  However, given their large
mass, such sleptons may be non-relativistic.  They can therefore be
highly ionizing, allowing one to distinguish them from genuine muons.
In addition, time-of-flight information could be used to detect a slow
moving particle.  Such signals are almost background free, providing
the potential for a spectacular
signature~\cite{Drees:1990yw,Goity:1993ih,Nisati:1997gb}.

Searches for metastable sleptons have been motivated previously by the
existence of such particles in gauge-mediated supersymmetry breaking
models~\cite{Feng:1997zr,Nisati:1997gb,Ambrosanio:1997,%
Mercadante:2000hw,Ambrosanio:2000}.  No signals have been found at
Tevatron Runs I and II~\cite{champsearches} and LEP~\cite{leptrack}.
The most stringent current bound is $m_{\tilde{l}} > 99~\gev$ at
95$\%$ CL from LEP searches at center-of-mass energies up to 208 GeV.
The prospects for a full Tevatron Run II have been investigated in
Ref.~\cite{Feng:1997zr}, where appropriate cuts in slepton velocity
and pseudorapidity $\eta$ have been included to eliminate the
background.  Requiring 5 or more events for a signal, the estimated
reach in right handed slepton mass is about 110 GeV, 180 GeV, and 230
GeV for integrated luminosities of 2, 10, and $30~\ifb$, respectively.
The discovery reach of the LHC has also been
considered~\cite{LHCstable}.  For one year at the design luminosity of
$100~\ifb$, metastable sleptons with mass up to 700 GeV could be
discovered.

As noted above, metastable sleptons are possible in both the
high-scale supersymmetry breaking scenarios discussed here, and in
low-scale supersymmetry breaking models, such as those with
gauge-mediated supersymmetry breaking.  In the gauge-mediated
scenarios, however, the gravitino mass is much lighter, around the keV
scale.  It is possible that these cases may be distinguished
cosmologically. Alternatively, direct collider searches for other
supersymmetric particles and the measurement of their mass spectra
will provide additional means for distinguishing these
possibilities. Finally, the slepton lifetimes in gauge-mediated
models, although long on collider detector scales, are much shorter
than in the superWIMP scenarios discussed here, and this may be
distinguished experimentally, providing an unambiguous
determination~\cite{FMS}.

\subsection{Sneutrino and Neutralino NLSP}

The cases of a gravitino LSP with either a sneutrino or neutralino
NLSP are qualitatively different from the slepton NLSP case.  The
allowed regions of parameter space are given in
\figsref{summary_sneu}{summary_Bino}.  In both cases, the metastable
NLSP will pass through detectors, resulting in missing energy
signatures topologically identical to the conventional missing energy
signal of supersymmetry.  There are four cases to distinguish: the
lightest standard model superpartner may be either a sneutrino or a
neutralino, and this particle may either decay to a gravitino or not.

\begin{figure}
\resizebox{3.25 in}{!}{
\includegraphics{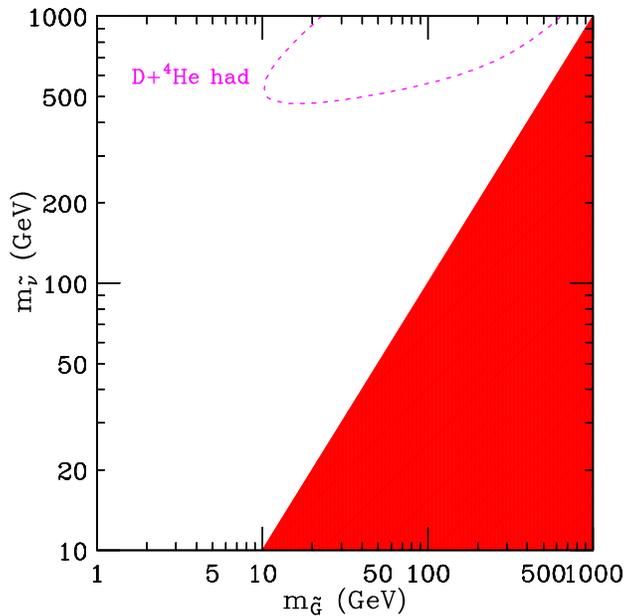}}
\caption{Excluded and allowed regions for the gravitino LSP scenario,
assuming a $\tilde{\nu}$ NLSP, as in \figref{sneu}, but now in the
$(m_{\gravitino}, m_{\tilde{\nu}})$ plane.  In the dark (red) shaded
region the gravitino is not the LSP.  The D $+$ $^4$He contour is as
in \figref{sneu}.
\label{fig:summary_sneu} }
\end{figure}

\begin{figure}
\resizebox{6.5 in}{!}{
\includegraphics{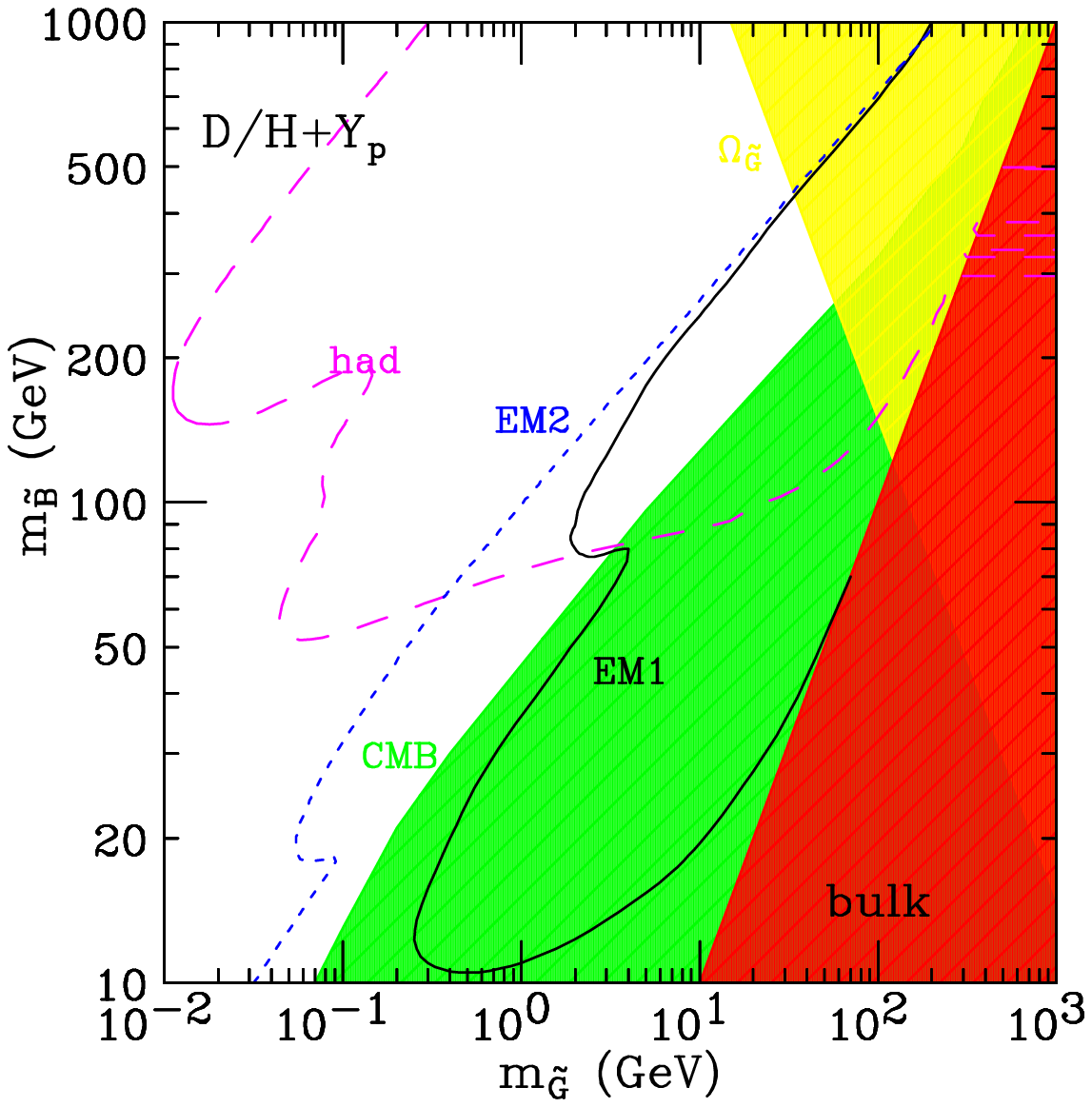}
\includegraphics{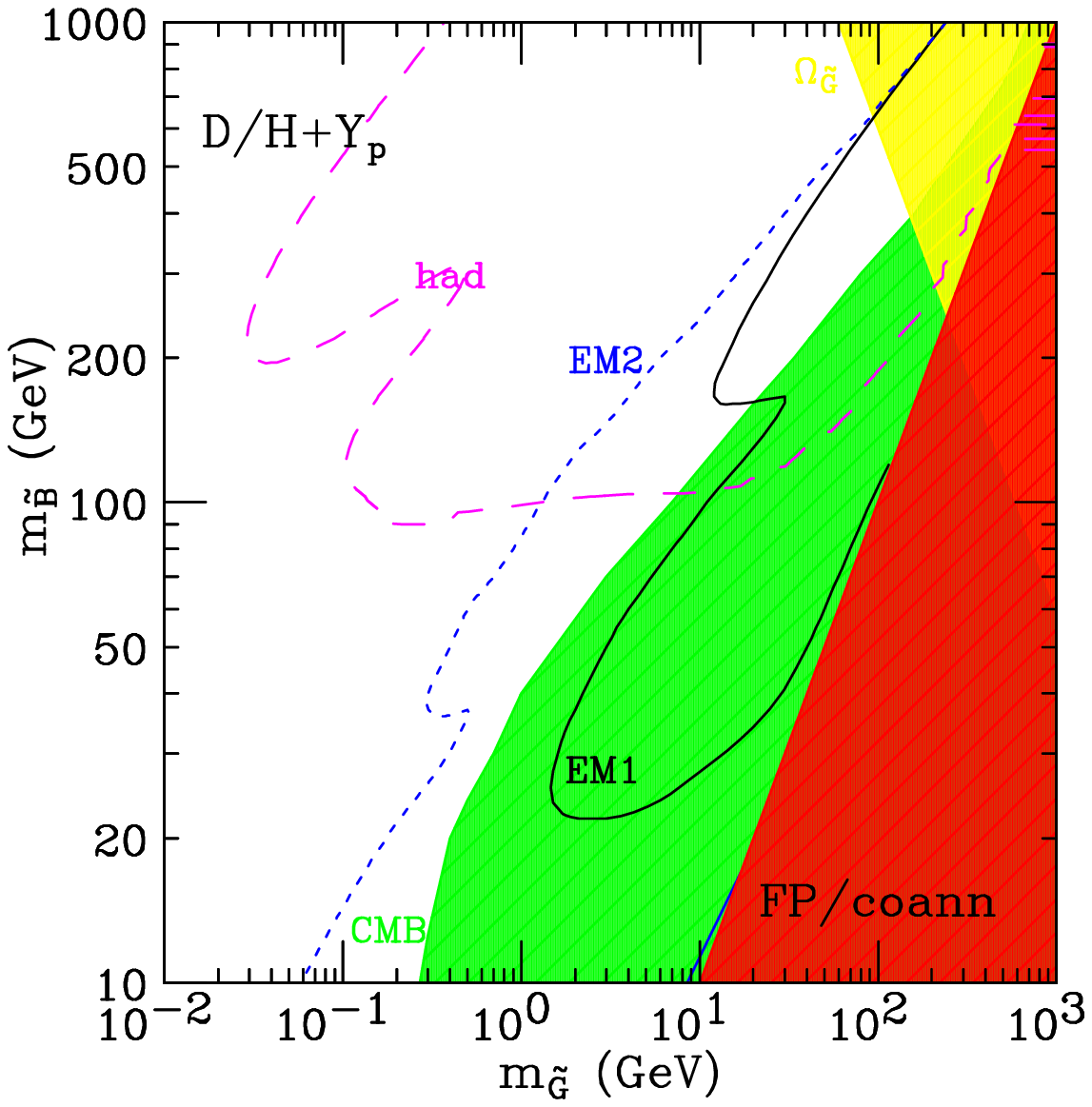}}
\caption{Excluded and allowed regions for the gravitino LSP scenario,
assuming a $\tilde{B}$ NLSP in the $(m_{\gravitino}, m_{\tilde{B}})$
plane.  In the (dark) red shaded region the gravitino is not the LSP.
All other shaded regions and contours are as in the left panel of
\figref{Bino_bulk} (left) and as in the left panel of \figref{Bino_fp}
(right).
\label{fig:summary_Bino} }
\end{figure}

The sneutrino and neutralino cases may be distinguished by precision
supersymmetry studies at colliders.  For example, in $e^+e^-$
collisions, the signatures of slepton pair production in the sneutrino
scenario (for example, $e^+ e^- \to \tilde{l} \tilde{l} \to l \nu
\tilde{\nu} q \bar{q}' \tilde{\nu}$) are identical to the signatures
of chargino production in the neutralino scenario (for example, $e^+
e^- \to \tilde{\chi}^+ \tilde{\chi}^- \to l \nu \tilde{\chi}^0 q
\bar{q}' \tilde{\chi}^0$)~\cite{deGouvea:1998yp}.  However, these
possibilities may be distinguished easily through angular
distributions at a linear collider, and possibly also at the LHC.

Determining whether the sneutrino or neutralino eventually decays to a
gravitino may be more difficult.  In the sneutrino case, the working
assumption would be that the gravitino is the LSP --- the sneutrino
itself is disfavored as a dark matter candidate, because, in the
natural region of parameter space, it predicts dark matter signals
that have not been seen.  On the other hand, in the neutralino case,
the working assumption would be that the neutralino is stable --- as
we have seen, decays to the gravitino are highly constrained by
hadronic BBN bounds.  Further information will be provided by high
sensitivity collider and astrophysical experiments.  For example, a
positive detection in dark matter search experiments would eliminate
the possibility of a gravitino LSP.  On the other hand, the precise
determination of supersymmetry parameters will make possible the
determination of the thermal relic abundance of the lightest standard
model superpartner.  This will favor the gravitino LSP scenario if
this thermal relic density is larger than the observed dark matter
density.

The current experimental limit on the sneutrino mass, assuming three
degenerate families, is $m_{\tilde{\nu}}>44.6~\gev$ at 95\% CL from
limits on the invisible decay width of the $Z$~\cite{Hebbeker:1999pi}.
The metastable sneutrino scenario has not been well-studied.  The
collider signals and detector reaches depend crucially on the identity
of the second lightest sparticle (which decays into the sneutrino),
the accompanying decay products, experimental cuts and details of the
detector.  A detailed study of the collider phenomenology of the
sneutrino scenario will appear in a future work. 

\section{Summary and Outlook}

In this paper, we have determined the viability of supergravity
scenarios in which the gravitino is the LSP.  We have considered the
three possibilities in which the NLSP is a slepton, a sneutrino, and a
neutralino.  In each case, we determined the branching fractions of
the leading two- and three-body decays, and applied constraints from
the dark matter density, CMB, and both EM and hadronic BBN bounds.  We
found that the hadronic BBN constraints, previously neglected, are
extremely important, providing the most stringent limits in natural
regions of parameter space.

The gravitino LSP scenario opens up many connections between particle
physics and cosmology.  Consider, for example, the slepton NLSP
scenario.  At colliders, it may be possible to collect the less
energetic metastable sleptons in a detector and monitor this detector
for slepton decays.  By measuring the decay time distribution and the
energy of each produced lepton, one could independently determine both
the gravitino mass and the reduced Planck mass
$\mstar$~\cite{Buchmuller:2004rq,Feng:2004gn}.

A measurement of the gravitino mass determines the scale of
supersymmetry breaking, with implications for model building and dark
energy.  At the same time, such a measurement would determine a
particular place in the $(\mgravitino, m_{\text{NLSP}})$ plane.
Colliders would therefore shed light on the possible role of such new
physics in BBN, and more generally in the thermal history of the
Universe after NLSP freeze out at temperatures of around 10 GeV.  For
example, the $\mu$ parameter sensitivity of \eqref{mu} may be improved
by two orders of magnitude in the future by the DIMES
mission~\cite{DIMES}.  Such improvement would extend the CMB
sensitivity contour significantly.  The measurement of a $\mu$
distortion consistent with the determination of $(\mgravitino,
m_{\text{NLSP}})$ would provide a striking confirmation of the
underlying gravitino scenario.

The measurement of $\mstar$ would provide a precision test of the
supersymmetry predictions relating the properties of gravitinos to
those of gravitons, and also provide the first direct measurement of
the Planck scale on microscopic
scales~\cite{Buchmuller:2004rq,Feng:2004gn}.  The crucial question is
the feasibility of collecting a sizable sample of metastable NLSPs.
There has been an earlier study on the collection of very long lived
heavy charged leptons~\cite{Goity:1993ih}.  A more detailed analysis
of the trapping of sleptons at future colliders is also now under
study~\cite{FMS}.

\begin{acknowledgments}
We thank K.~Kohri, T.~Moroi, Y.~Santoso and V.~Spanos for helpful
correspondence. FT thanks Y.~Santoso for useful discussions about the
stau thermal relic abundance in supergravity models.  The work of JLF
was supported in part by National Science Foundation CAREER Award
PHY--0239817, and in part by the Alfred P.~Sloan Foundation.
\end{acknowledgments}


\end{document}

